\documentclass[a4paper,fleqn]{cas-dc}
\usepackage[numbers,sort&compress]{natbib}
\usepackage{graphicx}
\usepackage{amsmath,amssymb}
\usepackage{dcolumn}
\usepackage{color}
\usepackage{bm}

\newcommand{\e}{\mathrm{e}}
\newcommand{\ud}{\,\mathrm{d}}
\newcommand{\im}{\mathrm{i}}
\newcommand{\h}{\mathcal{H}}
\renewcommand{\Im}{\mathrm{Im}\,}

\begin{document}
\let\WriteBookmarks\relax
\def\floatpagepagefraction{1}
\def\textpagefraction{.001}
\shorttitle{Superconductor-topological insulator dual proximity effect}
\shortauthors{N.~Sedlmayr and A.~Levchenko}

 \title[mode = title]{Hybridization mechanism of the dual proximity effect in superconductor-topological insulator interfaces}

\author[1]{Nicholas Sedlmayr}[orcid=0000-0001-7511-2910]
\cormark[1]
\ead{sedlmayr@umcs.pl}
\ead[URL]{nick.sedlmayr.co.uk}
\address[1]{Institute of Physics, M.~Curie-Sk{\l}odowska University, 20-031 Lublin, Poland}

\author[2]{Alex Levchenko}[orcid=0000-0002-0319-386X]
\ead{levchenko@physics.wisc.edu}
\address[2]{Department of Physics, University of Wisconsin-Madison, Madison, Wisconsin 53706, USA}

\cortext[1]{Corresponding author}

\begin{abstract}
In this communication we consider generalities of the proximity effect in a contact between a conventional $s$-wave superconductor (S) nano-island and a thin film of a topological insulator (TI). A local hybridization coupling mechanism is considered and a corresponding model is corroborated that captures not only the induced unconventional superconductivity in a TI, but also predicts the spreading of topologically protected surface states into the superconducting over-layer. This dual nature of the proximity effect leads specifically to a modified description of topological superconductivity in these systems. Experimentally accessible signatures of this phenomenon are discussed in the context of scanning tunneling microscopy measurements. For this purpose an effective density of states is computed in both the superconductor and topological insulator. As a guiding example, practical applications are made for Nb islands deposited on a surface of Bi$_2$Se$_3$. The obtained results are general and can be applied beyond the particular material system used. Possible implications of these results to proximity circuits and hybrid hardware devices for quantum computation processing are discussed.
\end{abstract}


\maketitle

\section{Introduction}

The ongoing efforts in developing superconducting hardware that enables topologically protected fault-tolerant quantum computation \cite{Kitaev,Nayak} relies heavily on the physics associated with the proximity effect (PE) in Josephson junctions (JJ) and other related weak-links, see Ref.~\cite{Golubov} for a review and references therein. This general phenomenon occurs at the interfaces between different materials where superconducting correlations can be effectively induced into a nominally non-superconducting material over a distance of the order parameter coherence length. The required building block of topological protection is based on the elusive excitations that satisfy non-Abelian exchange statistics such as Majorana zero modes (MZMs) \cite{Beenakker}. At present there are two leading candidate platforms that host these quantum states and can be integrated into suitable qubit circuit designs. First, is the hybrid superconductor-semiconductor (S-Sm) nanowire junction where an interplay of strong Rashba spin-orbit coupling and conventional superconductivity leads to an effective $p$-wave pairing state that supports formation of MZMs in the proper regime of control parameters such as Zeeman splitting, controlled by a magnetic field, and chemical potential, that can be tuned by a gate voltage \cite{Lutchyn,Oreg}. A similar version of this structure with a two-dimensional electron gas realizing topological superconductivity in a planar Josephson junction has also been proposed \cite{Pientka}. Second, is the combination of a conventional superconductor (S) and topological insulator (TI) where, similarly, the interplay of induced pairing correlations and strong spin-momentum locking of gapless TI surface excitations results in a phase resembling a spinless $p$-wave superconductor, that does not break time reversal symmetry. This state supports Majorana bound states that can be nucleated at vortices \cite{Ivanov,Fu-Kane,Hughes,Ioselevich}. Apart from hybrid systems, one should add to the list intrinsic topological superconductors (TSC) \cite{Qi-Zhang} that offer some of the same functionality concerning the MZMs physics.  A planar two-dimensional geometry of the S-TI and TSC systems offers great technical advantages for: (i) scalability, as the number of vortices with MZMs scales proportionally to the sample area, (ii) proposed MZMs interferometry \cite{Vishveshwara,Stern}, and (iii) ultimately the opportunity to perform braiding operations with seemingly realistic protocols \cite{Hegde}.

\section{Hybrid topological hardware platforms}

The proximity effect and Josephson physics in hybrid topological systems, including various geometrical configurations and combinations of unconventional superconducting materials, have been extensively studied theoretically. Several designs of proposed circuits were utilized experimentally in functional devices. We briefly overview and highlight recent work as well as the outstanding challenges and open questions in the context of physics properties and efforts geared towards topological qubit development. The following presentation is organized by the material platforms.

\textit{(i) S-Sm platform}. Since the original measurements reported in Ref.~\cite{Mourik}, key signatures of MZMs in superconducting semiconductor (S-Sm) nanowires have been heavily scrutinized in the subsequent detailed experiments \cite{Das,Deng,Albrecht,Suominen,Gul,Zhang}. Major progress has been made in improving the quality of these devices and hard gap engineering under the proximity effect. A particularly important step forward for computing purposes is the implementation of a nanowire based hybrid transmon like qubit (gatemon) that has been validated \cite{Larsen,Casparis} and characterized for various configurations in the non-topological regime \cite{Vavilov}. However, conclusive evidence for a functional Majorana qubit remains to be demonstrated. One of the major obstacles in this path is that the desired topological phase in a qubit configuration requires the fine-tuned calibration of the chemical potential with the simultaneous precise control of a high magnetic field, which is notoriously difficult to achieve. Another issue is that defect decoherence, or disorder in general, and multiband effects may result in signatures that could be falsely interpreted as Majorana states \cite{Chen,Yu,Aguado1,Aguado2,Pan}. We note that while most of the experimental efforts have been focused on S-Sm hybrids based on In(Ga)As and InSb materials, gate-controlled induced superconductivity was recently realized in Ge based quantum well heterostructures and a Josephson field effect transistor was achieved \cite{Hendrickx,Vigneau} thus providing novel avenues for the search of MZMs.

\textit{(ii) S-TI and TSC platforms}. The prospect of incorporating S-TI-S junctions as inductive elements for topological qubits attracted great interest. The proposed $p$-wave symmetry of the induced superconductivity into TI surface states in these structures was tested in numerous experiments by phase-sensitive techniques such as measuring surface supercurrents in the Josephson effect \cite{Brinkman,Goldhaber-Gordon,Molenkamp,LiLu,Mason,Oostinga,Kurter-1,Kurter-2,Stehno,Sochnikov,Williams,Rokhinson}. Perhaps the most heavily studied proximitized TI materials to date are Bi$_2$Se$_3$, Bi$_2$Te$_3$ and HgTe, while interestingly doping the parent compound by either Cu$_x$ or Nb$_x$ reveals native topological superconductivity in Bi$_2$Se$_3$ \cite{Ando-TSC,Finck-TSC}. The coexistence of superconducting and topological orders in the regime of the proximity effect has been further tested \cite{S-TI-Coexistence-1,S-TI-Coexistence-2,S-TI-Coexistence-3}, including junctions of TI and unconventional $d$-wave superconductors \cite{Burch}, and a large volume of transport data is available, especially on tunneling conductance and Andreev reflection \cite{Finck-1,Tikhonov,Finck-2,Banerjee}. The key signature of MZMs in S-TI-S junctions, namely $4\pi$ periodic Josephson current, $I_J(\phi)\propto\sin(\phi/2)$, has not been so far unambiguously detected. However, field induced localization of MZMs in vortex cores and the spatial profile of the Majorana mode have been mapped out by scanning tunneling microscopy in epitaxial thin films of Bi$_2$Te$_3$-NbSe$_2$ heterostructures \cite{MFs-Vortex-BiTeNbSe}. In addition, robust zero-bias conductance peaks associated with MZMs have been recently identified by STM in iron-based superconductors \cite{MFs-Vortex-FeTeSe,MFs-Vortex-LiFe,MFs-Vortex-FeSeTe}. These topological materials, and FeTe$_{1-x}$Se$_x$ in particular, have been recognized as a promising new candidate for further study and possible integration into circuits for computing applications.

\textit{(iii) CPB and JJA platforms}. The alternative ideas in superconducting electronics to realize quantum states robust against sources of local noise and bit flips are also being actively pursued where the notion of topological protection is encoded in a different system property. The advantages of these approaches is that they employ only conventional materials and mature technologies based on Cooper-pair boxes (CPB) and Josephson junction arrays (JJA). The context of topological protection in these circuits can be introduced in a different way by encoding the state of the qubit by a discrete variable that cannot be measured without taking the qubit out of the protected domain. For instance, the simplest example involves two logical states constructed as coherent superpositions of a large number of states with e.g.~even or odd quantum numbers. Clearly, such states are physically indistinguishable and thus protected from the environment, provided that transitions between the odd and even sectors are mitigated. The promising implementation of this idea is to identify such discrete quantum states with dual variables of either charge or phase parity representing the number of Cooper pairs on a small superconducting island or the number of trapped fluxons, respectively. This architecture requires only conventional materials such as Al-AlO$_x$-Al junctions with an artificially engineered $\pi$-periodicity of Josephson energy, $E_J(\phi)\propto\cos(2\phi)$. Physically, two different superconducting circuits have been suggested to implement such charge parity qubits: the $0-\pi$ mirror qubit \cite{Kitaev-SC-mirror-Qubit,Preskill,Koch} and the rhombi chains \cite{Doucot-1,Doucot-2}, recently realized experimentally \cite{Schuster,Gershenson-CPQ-NP,Gershenson-CPQ-PRL,Marcus-CPQ}. Phase dual circuits based on Aharonov-Casher effect have also been studied \cite{Gershenson-PPQ,Bell-Bifluxon}.

\textit{(iv) Multiterminal JJ platforms}. Another fruitful method tailored towards achieving topological properties with conventional materials explores the potential of multiterminal proximity circuits as artificial emulators of topological matter \cite{Nazarov}. In this approach when the normal region of a Josephson junction is contacted by $n$ superconducting leads the spectra of emergent Andreev levels are consequently parametrized by $(n-1)$ phases. Since the spectrum of levels is periodic modulo $2\pi$, the pairwise difference of phases between each of the two terminals can be thought as an effective dimension in reciprocal momentum space and thus Andreev bands emulate an effective band-structure of a crystal that lives in such a parameter space of phases. Remarkably, under certain experimentally realistic conditions, the Berry flux associated with these sheets of Andreev states can be made finite as the band structure supports stable Dirac (and more generally Weyl) points \cite{Meyer,Xie-1}. These features lead to topological properties such as non-local conductance and current of Cooper pair pumps that are asymptotically quantized \cite{Eriksson,Xie-2,Leone} (the accuracy of these quantizations is determined by dissipative Landau-Zener transitions between Andreev states). Provided that current superconducting qubit circuits contain multiple leads, realizing and operating such a multiterminal Andreev qubit in a parameter space of the topological domain is of clear interest. Incorporating topological materials into such JJs further enriches the spectrum of possibilities \cite{Houzet,Xie-3}. In the past few years the multiterminal Josephson effect was experimentally realized in proximitized graphene \cite{Finkelstein} and InAs/Al epitaxial heterostructures \cite{Pribiag,Manucharyan}.

In the context of qubit performance, the fundamental issue that pertains to each of the above hybrid platforms is the quasiparticle poisoning that severely limits qubit lifetimes. The combination of disorder effects and magnetic fields, which break time reversal symmetry, contributes to the softening of the proximity gap and induces tails of sub-gap states thus enabling quasiparticle population down to the lowest energies. Experimentally, gap engineering, with a combination of materials in proximity effects with large gap mismatch, was effectively used to stimulate quasiparticle relaxation and trapping. Theoretically, the sub-gap structure in the density of states was exhaustively studied, especially in S-Sm structures. The same set of issues is expected to be a limiting factor in the performance of the Majorana qubit in S-TI platforms, but far fewer detailed studies are available for this system that go beyond the basic models. This motivates the current work where we formulate and solve the hybridization model of the proximity effect in S-TI interfaces, and discuss its dual nature in Sec.~\ref{Sec:Top-PE}. We extract effective density of states in SC and TI in Sec. \ref{sec_dos} and develop effective models in Sec.~\ref{sec_eff}. We specify our analysis to a particular geometry amenable to surface state measurement instruments such as scanning tunneling and magnetic force microscopes that can test predictions of the extended model. We close with a brief summary and outlook in Sec.~\ref{Sec:Summary}.

\begin{figure}[t!]
\centering
\includegraphics[width=0.99\columnwidth]{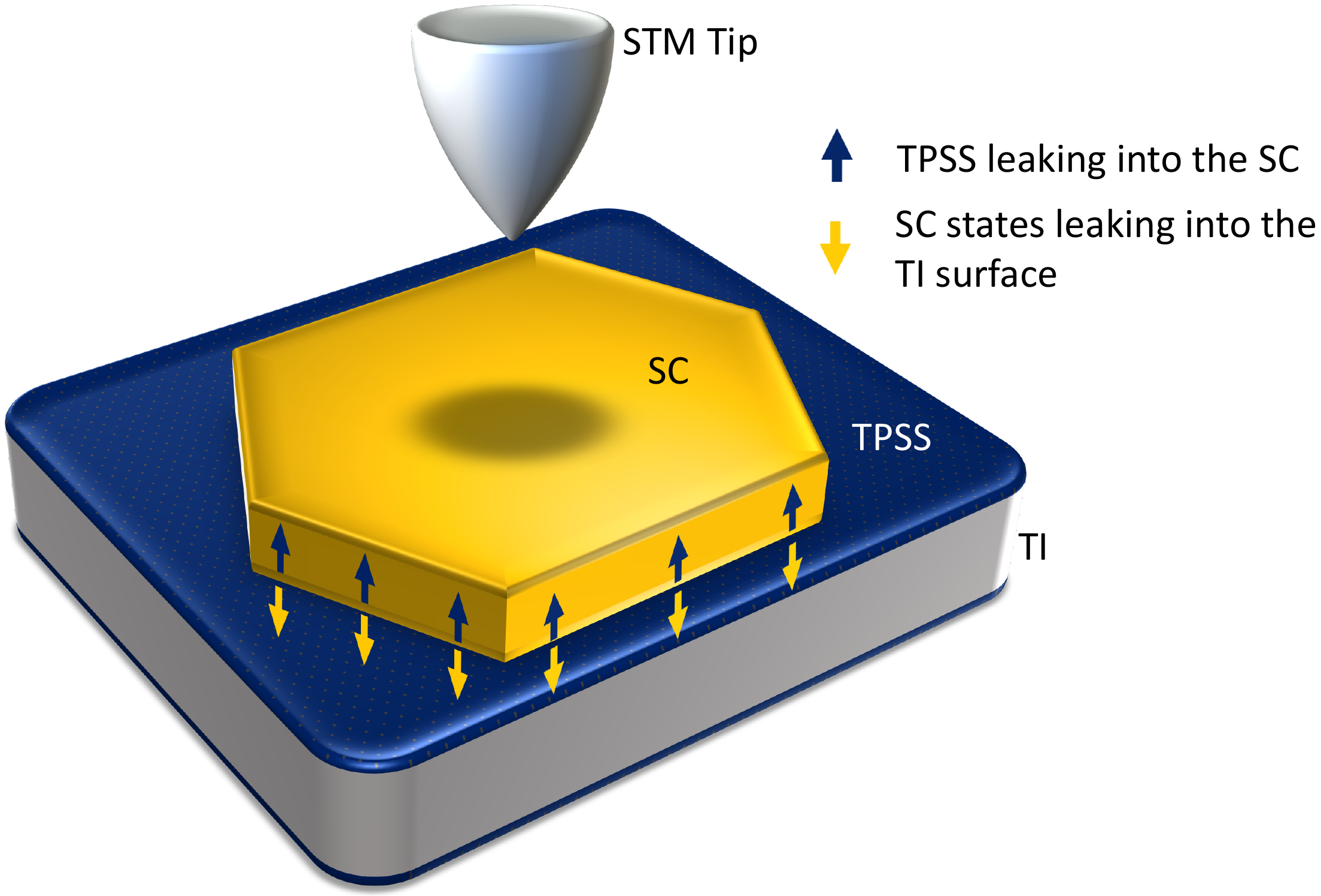}
\caption{A schematic illustration of the set-up considered for the model. A SC island is on top of a bulk 3D topological insulator, with TPSS on its top surface. STM experiments allow one to probe the density of states on both the surfaces of the SC island and TI.}
\label{ref:figure_schematic}
\end{figure}

\section{Hybridization model for the topological PE}\label{Sec:Top-PE}

In this work we focus on the frequently overlooked dual nature of the proximity effect in S-TI layers. The term topological PE is used to describe the generically expected phenomenon of topologically protected surface states (TPSS) leaking into the adjacent superconducting material, thus effectively inducing properties native to intrinsic TSCs. This effect opens new perspectives for observing and manipulating MZMs.

From the recent modeling there exists a substantial supporting evidence that TPSS can migrate and spread in multilayer heterostructures \cite{Wu-Zhang}. There are two kinds of experiments that support this conclusion. First are the ARPES measurements \cite{Trang} that reveal that a Dirac cone of TPSS can be seen on the surface of superconducting film grown on top of TI. This effect was observed in Pb-TlBiSe$_2$ proximity layers. Second are STM data \cite{Dayton,Sedlmayr} also suggesting clear signatures of Dirac states in SC density of states spectra. The theoretical investigation of these features have been carried out in both S-TI systems \cite{Lababidi,Sedlmayr,Hugdal,Bobkova} and with numerical simulations in topological superconducting wires \citep{Guigou,Ptok}.

In this study we expand on our earlier proposal, reported in conjunction with the experiment \cite{Sedlmayr}, and demonstrate with a more detailed modeling that hybridization between the TPSS and superconductor results in effective $p$-wave like topological superconductor with striking profiles in the density of states. In particular, we discuss how the Fu and Kane model \cite{Fu-Kane} of proximity induced pairing can be recovered in the proper limit of the present treatment, as induced superconductivity of the TPSS is not completely equivalent to the latter mechanism.
As a practical application, motivated by geometries accessible to surface probes, we investigate the possible experimental setup of a superconducting droplet deposited on top of a TI, see Fig.~\ref{ref:figure_schematic} for a schematic of the system. In the subsequent analysis we assume that the interface is clean and contact between the two materials is without potential barriers. For estimates we take a combination of Nb islands as SC and paradigmatic Bi$_2$Se$_3$ as TI. The minimal Hamiltonian we use which captures this physics has three terms:
\begin{equation}\label{ham1}
H=H_{\rm S}+H_{\rm TI}+H_{\rm C}\,.
\end{equation}
$H_{\rm S}$ is the standard BCS Hamiltonian for a clean two dimensional metal with mean field $s$-wave pairing. $H_{\rm TI}$ describes the TI surface states (TPSS), and $H_{\rm C}$ is a spin independent local hybridization between the TPSS and the SC states. These are all described in more detail below.

Throughout we will use $H_{\rm j}$ for the Hamiltonian, with the corresponding Hamiltonian density $\h_{\rm j}$ such that for example
\begin{equation}
 H_{\rm S}=\int\ud^2{\mathbf{r}}\Psi^\dagger_{\mathbf{r}}\h_{\rm S}\Psi_{\mathbf{r}}\,.
\end{equation}
${\mathbf{r}}=(x,y)$ is the 2D spatial coordinate. For the SC states we use the Nambu basis, with $\Psi^\dagger_{\mathbf{r}}=\{c^\dagger_{{\mathbf{r}}\uparrow},c^\dagger_{{\mathbf{r}}\downarrow},c_{{\mathbf{r}}\downarrow},-c_{{\mathbf{r}}\uparrow}\}$, and a wavefunction $\psi^T_{\mathbf{r}}$: $\{u_{{\mathbf{r}}\uparrow},u_{{\mathbf{r}}\downarrow},v_{{\mathbf{r}}\downarrow},v_{{\mathbf{r}}\uparrow}\}$. $c^\dagger_{{\mathbf{r}}\sigma}$ creates a particle of spin $\sigma$ at position ${\mathbf{r}}$. Similarly for the TPSS we have $\chi^\dagger_{\mathbf{r}}=\{a^\dagger_{{\mathbf{r}}\uparrow},a^\dagger_{{\mathbf{r}}\downarrow},a_{{\mathbf{r}}\downarrow},-a_{{\mathbf{r}}\uparrow}\}$ and
\begin{equation}
H_{\rm TI}=\int\ud {\mathbf{r}}\chi^\dagger_{\mathbf{r}}\h_{\rm TI}\chi_{\mathbf{r}}\,,
\end{equation}
where $a^\dagger_{{\mathbf{r}}\sigma}$ creates a particle of spin $\sigma$ at position $\mathbf{r}$. Additionally we will set $\hbar=k_B=1$ throughout.

First, the BCS Hamiltonian density is given by
\begin{equation}\label{hamsc}
\h_{\rm S}=\hat{\xi}{\bm \tau}^z+\Delta{\bm \tau}^x\,.
\end{equation}
Note that we follow the convention that identity matrices are not explicitly written, so equation \eqref{hamsc} is a $4\times4$ matrix including the spin subspace.  As is standard $\vec{\bm\tau}$ are the Pauli matrices operating in the particle-hole subspace and $\vec{\bm\sigma}$ are the Pauli matrices operating in the spin subspace. The band operator is $\hat\xi=-\nabla^2/(2m)-\mu$, with $\mu$ the chemical potential and $m$ the effective electron mass.

For the the TPSS we have \cite{Zhang-Zhang}
\begin{equation}
\h_{\rm TI}=\left(-\im v_F\nabla\cdot\vec{\bm \sigma}-{\tilde{\mu}}\right){\bm \tau}^z\,.
\end{equation}
$\tilde{\mu}$ is the chemical potential of the TPSS, which in general can be different to $\mu$. Note that in our hybrid model the velocity of the TPSS, $v_F$, will depend on the thickness of the superconducting layer. A simple hydrodynamic consideration suggests that $v_F=v_{F0}\beta\ell_{\rm TI}/(\ell_{\rm TI}+\ell_{\rm S})$ where $\ell_{\rm TI}$ is the penetration depth of the TPSS into the TI bulk and $\ell_{\rm S}$ is the thickness of the SC layer. $v_{F0}$ is the Fermi velocity of the TPSS in the absence of the SC and the dimensionless parameter $\beta$ has been introduced to model any further effects of the interface itself.

For the hybridization we consider the simplest possible coupling term
\begin{equation}
H_{\rm C}=\gamma\int\ud^2{\mathbf{r}}\chi^\dagger_{\mathbf{r}}{\bm \tau}^z\Psi_{\mathbf{r}}+\textrm{H.c.}\,,
\end{equation}
with a strength $\gamma$. For our model we can also expect that the coupling decays exponentially with the thickness.

\begin{figure}[t!]
\centering
\includegraphics[width=0.95\columnwidth]{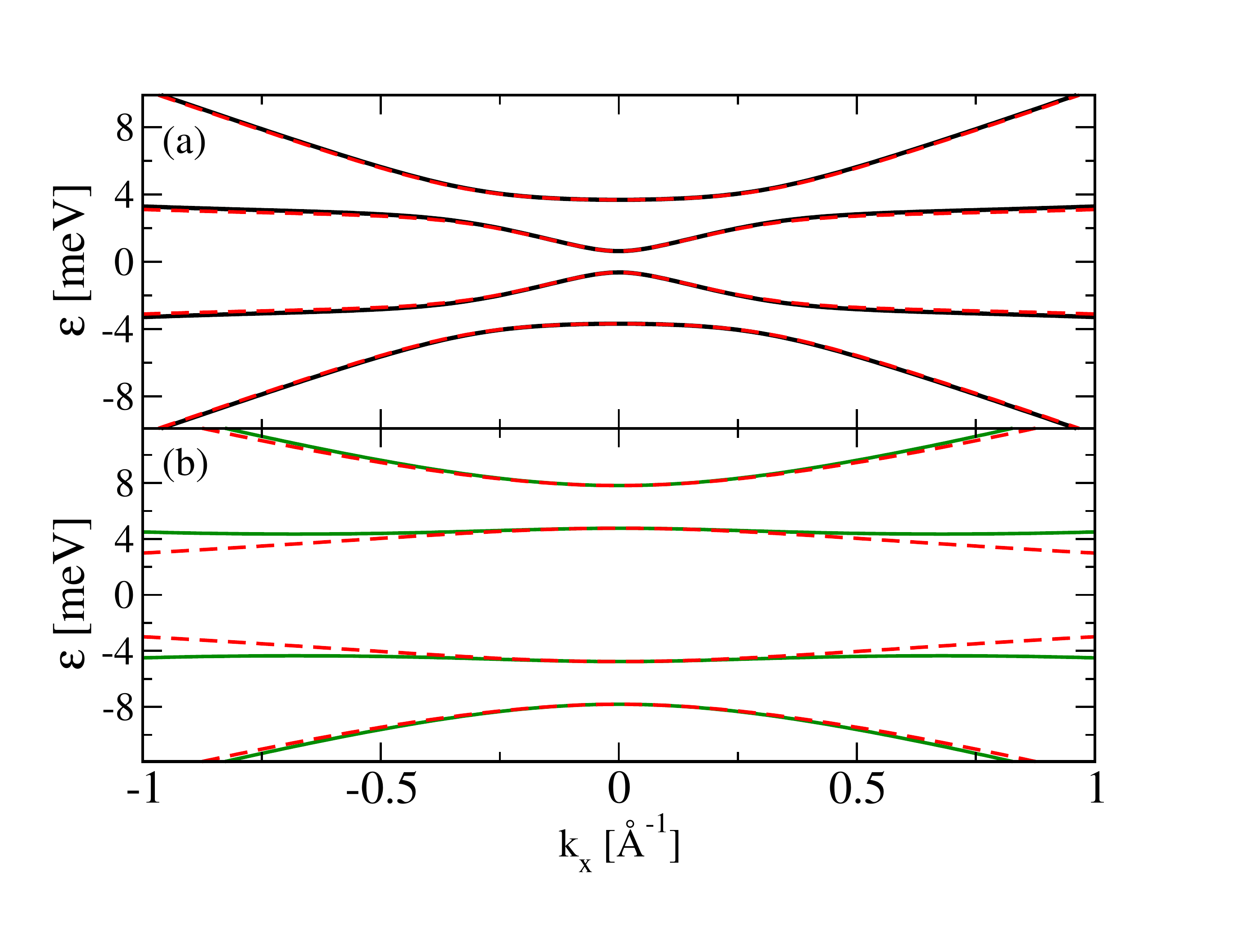}
\caption{Band structure for the hybridization between the approximately flat superconducting bands and the Dirac cone belonging to the TPSS is clearly visible in (a). In both panels $v_F=10$ meV{\AA}. The coupling is in (a) $\gamma=0.5\Delta$ and in (b) $\gamma=2\Delta$.}
\label{ref:figure_band}
\end{figure}

The Hamiltonian \eqref{ham1} can be decoupled into two independent blocks by a Fourier transform, followed by a spin rotation which diagonalizes $\h_{\rm TI}$ but naturally leaves $\h_{\rm S}$ and $\h_{\rm C}$ unaffected. Following the Fourier transform the Hamiltonian densities are diagonal in momentum space with momentum $\mathbf{k}=(k_x,k_y)$ and $k=|\bf{k}|$. Now let $\widetilde\Psi_{\bf{k}}=\mathcal{V}_{\bf{k}}^\dagger\Psi_{\bf{k}}$ and $\widetilde\chi_{\bf{k}}=\mathcal{V}_{\bf{k}}^\dagger\chi_{\bf{k}}$, with $\widetilde\h=\mathcal{V}_{\bf{k}}^\dagger\h\mathcal{V}_{\bf{k}}$ where $\mathcal{V}_{\bf{k}}$ is a spin rotation which commutes with $\h_{\rm S}$ and $\h_{\rm c}$. Its exact form is given by
\begin{equation}
\mathcal{V}_{\bf{k}}=\e^{\im\phi_k{\bm \sigma}^z/2}
\e^{\im\pi{\bm \sigma}^x/4}\,,
\end{equation}
where $\phi_k=\pi/2-\tan^{-1}[k_x/k_y]$ is the polar angle coordinate. The rotated Hamiltonian density for the TPSS becomes
\begin{equation}
 	\widetilde\h_{\rm TI}=\mathcal{V}_{\bf{k}}^\dagger\h_{\rm TI}\mathcal{V}_{\bf{k}}=(v_Fk{\bm\sigma}^z-{\tilde{\mu}}){\bm \tau}^z\,.
\end{equation}
As stated previously $\widetilde\h_{\rm S}=\h_{\rm S}$ and $\widetilde\h_{\rm C}=\h_{\rm C}$.

In the new basis the Hamiltonian Eq.~\eqref{ham1} can be directly written in block diagonal form
\begin{equation}
    \widetilde\h\to\begin{pmatrix}
    \widetilde\h^+&0\\0&\widetilde\h^-
    \end{pmatrix}
\end{equation}
by a simple reordering. The resulting two independent $4\times4$ Hamiltonian densities are
\begin{equation}\label{ham2}
\widetilde\h^\pm=
\begin{pmatrix}
\xi_k&\gamma&\Delta&0\\
\gamma&\zeta_{\pm k}&0&0\\
\Delta&0&-\xi_k&-\gamma\\
0&0&-\gamma&-\zeta_{\pm k}
\end{pmatrix}\,,
\end{equation}
with $\zeta_k=v_Fk-{\tilde{\mu}}$, i.e.~the dispersion of the uncoupled TPSS. If we consider the anti-unitary symmetries used to classify topological Hamiltonians in the ten-fold way \cite{Schnyder} then strictly this Hamiltonian possesses $[\mathcal{C},\h^\pm]_+=0$ and $[\mathcal{T},\widetilde\h^\pm]_-=0$. These anti-unitary symmetries obey $\mathcal{C}^2=-1$ and $\mathcal{T}^2=1$ placing it in CI, which has no invariant in two dimensions. Nonetheless we can find effective models, see section \ref{sec_eff}, which are 2D topological superconductors with the possibility of hosting MZMs.

From Eq.~\eqref{ham2} it is straightforward to calculate the dispersion of the resulting energy bands:
\begin{align}\label{spectrum}
&\varepsilon^{abc}_k=\frac{a}{\sqrt{2}}
\left[b\left(
2\gamma^2+\epsilon_k^2+\zeta_{ck}^2\right)
\phantom{\sqrt{\left[\zeta_{ck}\right]}}\right. \nonumber \\
&+\left.\sqrt{\left[\epsilon_k^2-\zeta_{ck}^2\right]^2+4\gamma^2\left[\epsilon_k^2+2\xi_k\zeta_{ck}+\zeta_{ck}^2\right]}\right]^{\frac{1}{2}}\,,
\end{align}
where $a,b,c$ can each be either $\pm1$, and $\epsilon_k=\sqrt{\Delta^2+\xi_k^2}$. As required, in the limit $\gamma\to 0$ the BCS and TPSS dispersion relations are recovered. A full superconducting gap remains provided that the coupling $\gamma$ is strong enough. However for very small $\gamma$ there are mid-gap states due to the TPSS only having a weak pairing effect. In figure \ref{ref:figure_band} we show exemplary band structures cut across one direction of the 2D plane. As the dispersion relations are circularly symmetric this suffices to show all information. Throughout this paper we use the values $\mu={\tilde{\mu}}=0$, an effective mass of $m=3.2m_e$ where $m_e$ is the bare electron mass, and $\Delta=3.05$meV \cite{Sedlmayr}. In section \ref{sec_dos} we calculate the density of states for this model, and compare it to the standard BCS density of states and that for the Fu-Kane model, see section \ref{fkdos}.

\begin{figure}[t!]
\centering
\includegraphics[width=0.85\columnwidth]{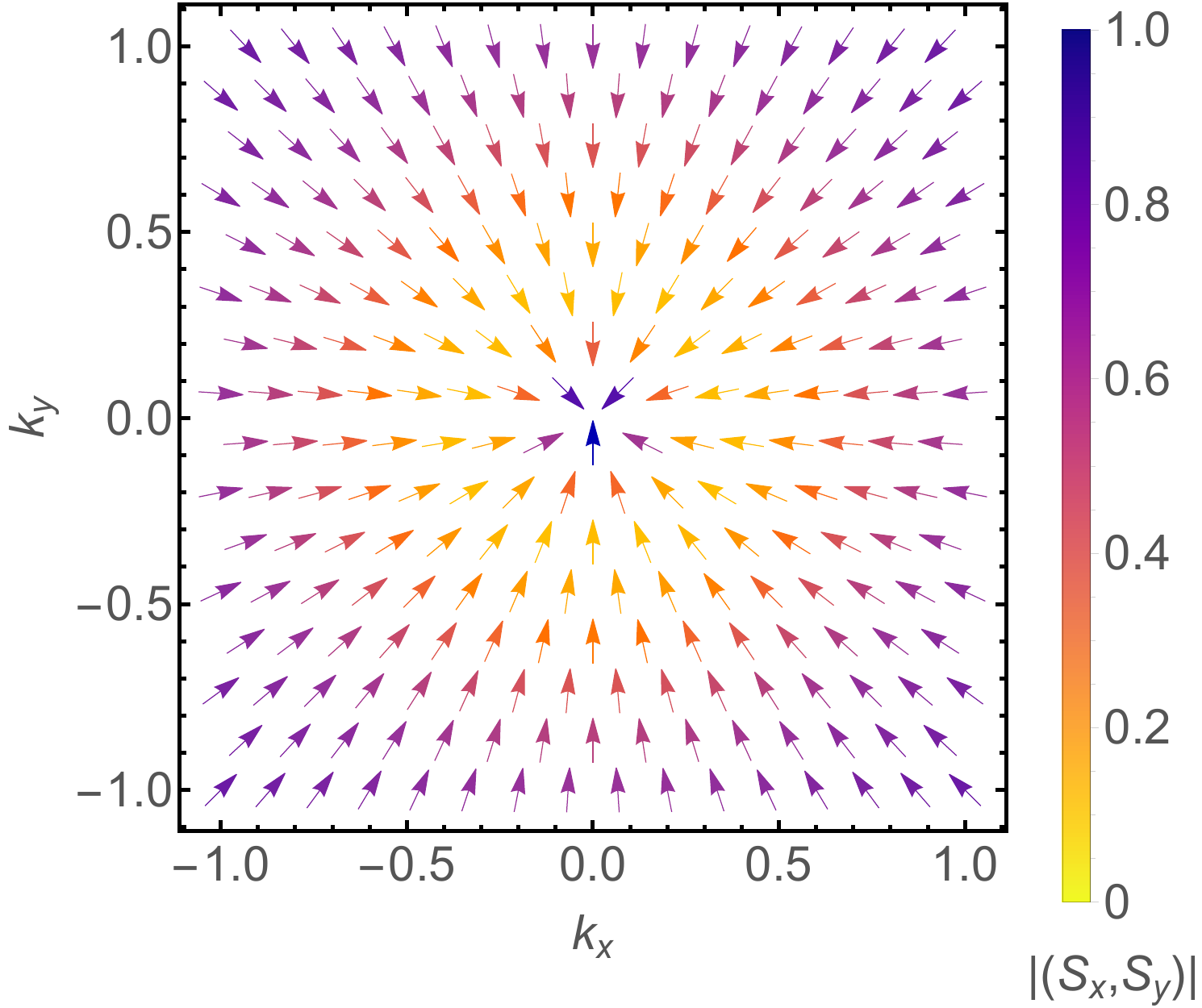}
\caption{The planar spin density $(S_x,S_y)$ as a function of momentum for the lowest positive energy band where $v_F=10$ meV{\AA} and $\gamma=0.5\Delta$, see figure \ref{ref:figure_ban d}(a). The spin momentum locking is clearly visible, and can be seen in all bands. There is also a component of spin in the $z$ direction, not shown.}
\label{ref:figure_spin}
\end{figure}

Another check one can perform that the hybridized bands have the right properties that can lead to topological superconductivity in the appropriate limit, is that there is spin momentum locking in the gapped Dirac cones near $k=0$. In figure \ref{ref:figure_spin} we show the in-plane spin density in momentum space, centred on $k=0$, for the lowest positive energy band. Spin-momentum locking is clearly visible, and a similar pattern can be seen for all the bands.

\section{Density of States}\label{sec_dos}

The density of states of the hybrid structure can be directly measured on the surface by scanning tunneling microscopy, see figure \ref{ref:figure_schematic}. In this section we will show that in the right regime the hybridization of the TPSS and the SC states leaves a clear measurable signature in the density of states.

For our full hybrid model, given the dispersion relation \eqref{spectrum}, one can directly calculate the density of states. The density of states is given by
\begin{equation}\label{fulldos}
\nu(\omega)=\frac{1}{\sqrt{\pi}\Gamma}\int\frac{\ud^2k}{4\pi^2}
\sum_{\{a,b,c\}} \e^{-(\omega-\varepsilon^{abc}_{\bf{k}})^2/\Gamma^2} \,,
\end{equation}
where the delta function peaks have been broadened into Gaussians of width $\Gamma=1$meV. This we then calculate numerically with Eq.~\eqref{spectrum}. To compare to this we consider the density of states of the Fu-Kane model, see section \ref{fkdos} below, and the standard BCS density of states given by
\begin{equation}\label{bcsdos}
\nu_{\rm S}(\omega)=\frac{1}{m\pi}\left|\textrm{Re}\frac{\omega}{\sqrt{\omega^2-\Delta^2}}\right|\,.
\end{equation}
For the purposes of plotting the density of states we introduce a phenomenological broadening via $\omega\to\omega+\textrm{i}\delta$.

\subsection{Fu-Kane model density of states}\label{fkdos}

As a comparison we can compare the density of states of our model to that of the Fu-Kane model. In the limit that momentum $k\to0$, i.e.~at the centre of the TPSS Dirac cone, our model \eqref{effh} can effectively be reduced to the Fu-Kane model, see section \ref{sec_eff_fk}. The Fu-Kane model can be succinctly written as \cite{Fu-Kane},
\begin{equation}\label{hfk}
\h_{\rm FK}=(v_Fk{\bm\sigma}^z-{\tilde{\mu}}){\bm \tau}^z+\Delta{\bm \tau}^x\,.
\end{equation}
An analytical expression for the density of states of this Hamiltonian can be calculated by starting from Gorkov's equation
\begin{equation}\label{gti}
\begin{pmatrix}
\im\omega_n-\h'_{\bm k} & {\bm \Delta}\\-{\bm \Delta}^\dagger & -\im\omega_n-\left[\h'_{-\bm k} \right]^*
\end{pmatrix}
\begin{pmatrix}
G_{n,{\bm k}}\\\mathcal{F}^\dagger_{n,{\bm k}}
\end{pmatrix}
=
\begin{pmatrix}
1\\0
\end{pmatrix}\,,
\end{equation}
with
\begin{align}
\h'_{\bm k}=&\begin{pmatrix}
-{\tilde{\mu}}&v_{F}k_-\\
v_{F}k_+&-{\tilde{\mu}}
\end{pmatrix}\,,
\end{align}
where $k_\pm=k_x\pm \im k_y$. The matrix ${\bm \Delta}=\Delta i{\bm \sigma}^y$ is the $s$-wave pairing. This is in the Matsubara representation with the Green's function
\begin{equation}\label{matg}
G_{n,{\bm k}}=-\int_0^\frac{1}{T}\ud\tau
\e^{\im \omega_n\tau}
\langle\mathcal{T}_\tau\psi_{{\bm k}\sigma}(\tau)\psi^\dagger_{{\bm k}\sigma'}(0)\rangle\,,
\end{equation}
and anomalous Green's function
\begin{equation}\label{matf}
\mathcal{F}^\dagger_{n,{\bm k}}=-\int_0^\frac{1}{T}\ud\tau
\e^{\im \omega_n\tau}
\langle\mathcal{T}_\tau\psi^\dagger_{{\bm k}\sigma}(\tau)\psi^\dagger_{{\bm k}\sigma'}(0)\rangle\,.
\end{equation}
The operator $\mathcal{T}_\tau$ stands for time ordering along the imaginary time axis and $\psi_{\sigma}({\mathbf{r}},\tau)$ is a Heisenberg operator. The Matsubara frequencies are $\omega_n=2\pi(n+\frac{1}{2})$ for $n\in\mathbb{Z}$, and temperature $T$.

The anomalous Green's function is therefore
\begin{equation}
\mathcal{F}^{\dagger}_{n,{\bm k}}=-\frac{\im\Delta{\bm \sigma}^y}{\varepsilon_{nk}^4-4{\tilde{\mu}}^2 v_F^2k^2}\begin{pmatrix}
\varepsilon_{nk}^2&-2{\tilde{\mu}} v_Fk_+\\
-2{\tilde{\mu}} v_Fk_-&\varepsilon_{nk}^2
\end{pmatrix}\,,
\end{equation}
where
\begin{equation}
\varepsilon_{nk}^2={\tilde{\mu}}^2+\omega_n^2+\Delta^2+v_F^2k^2\,.
\end{equation}
The normal Green's function can be found from the anomalous Green's function resulting in
\begin{equation}
G_{n,{\bm k}}=-\frac{\im\omega_n+\left[\h'_{-\bm k} \right]^*}{\varepsilon_{nk}^4-4{\tilde{\mu}}^2 v_F^2k^2}\begin{pmatrix}
\varepsilon_{nk}^2&-2{\tilde{\mu}} v_Fk_+\\
-2{\tilde{\mu}} v_Fk_-&\varepsilon_{nk}^2
\end{pmatrix}\,,
\end{equation}
from which the density of states can be calculated.

The Fu-Kane model density of states is calculated from the Green's function using the standard formula
\begin{equation}
\nu_{\rm FK}(\omega)=-\frac{1}{\pi}\Im \int\frac{d^2{\bm k}}{4\pi^2}\,\mathrm{tr} \left.G_{n,{\bm k}}\right|_{\im\omega_n\to\omega+\im\delta}\,,
\end{equation}
which after a small amount of manipulation leaves the integral
\begin{equation}
\nu_{\rm FK}(\omega)=\frac{2\omega}{\pi}\Im \int_{0}^\infty\frac{d k}{2\pi}\left.\frac{\varepsilon_{nk}^2k}{\varepsilon_{nk}^4-4{\tilde{\mu}}^2 v_F^2k^2}\right|_{\im\omega_n\to\omega+\im\delta}\,.
\end{equation}
This integral can be performed using standard techniques and we find,
\begin{align}\label{dosfk0}
\nu_{\rm FK}(\omega)=&\frac{|\omega|}{\pi v_F^2}\left[
\frac{\textrm{Re}\sqrt{\omega^2-\Delta^2-\tilde\mu^2}}{\sqrt{|\omega^2-\Delta^2-\tilde\mu^2}|}\right.\\\nonumber&\quad+\left.
\frac{|\tilde\mu|}{\sqrt{|\omega^2-\Delta^2|}}\frac{\textrm{Re}\sqrt{\Delta^2-\omega^2+\tilde\mu^2}}{\sqrt{|\omega^2-\Delta^2-\tilde\mu^2}|}
\right]\,,
\end{align}
and, in the case of interest $\tilde\mu=0$, this is simply
\begin{equation}\label{dosfk}
\nu_{\rm FK}(\omega)=\frac{|\omega|}{\pi v_F^2}
\Theta(\omega^2-\Delta^2)\,,
\end{equation}
where $\Theta$ is the Heaviside theta function. In the limit $\Delta\to0$ (and $\tilde\mu\to0$) one recovers the density of states,
\begin{equation}\label{dosdirac}
\nu_{\rm D}(\omega)=\frac{|\omega|}{\pi v_F^2}\,,
\end{equation}
for a Dirac cone in two dimensions.

\subsection{Density of states results}\label{sec:dos_results}

\begin{figure}[t!]
\centering
\includegraphics[width=0.99\columnwidth]{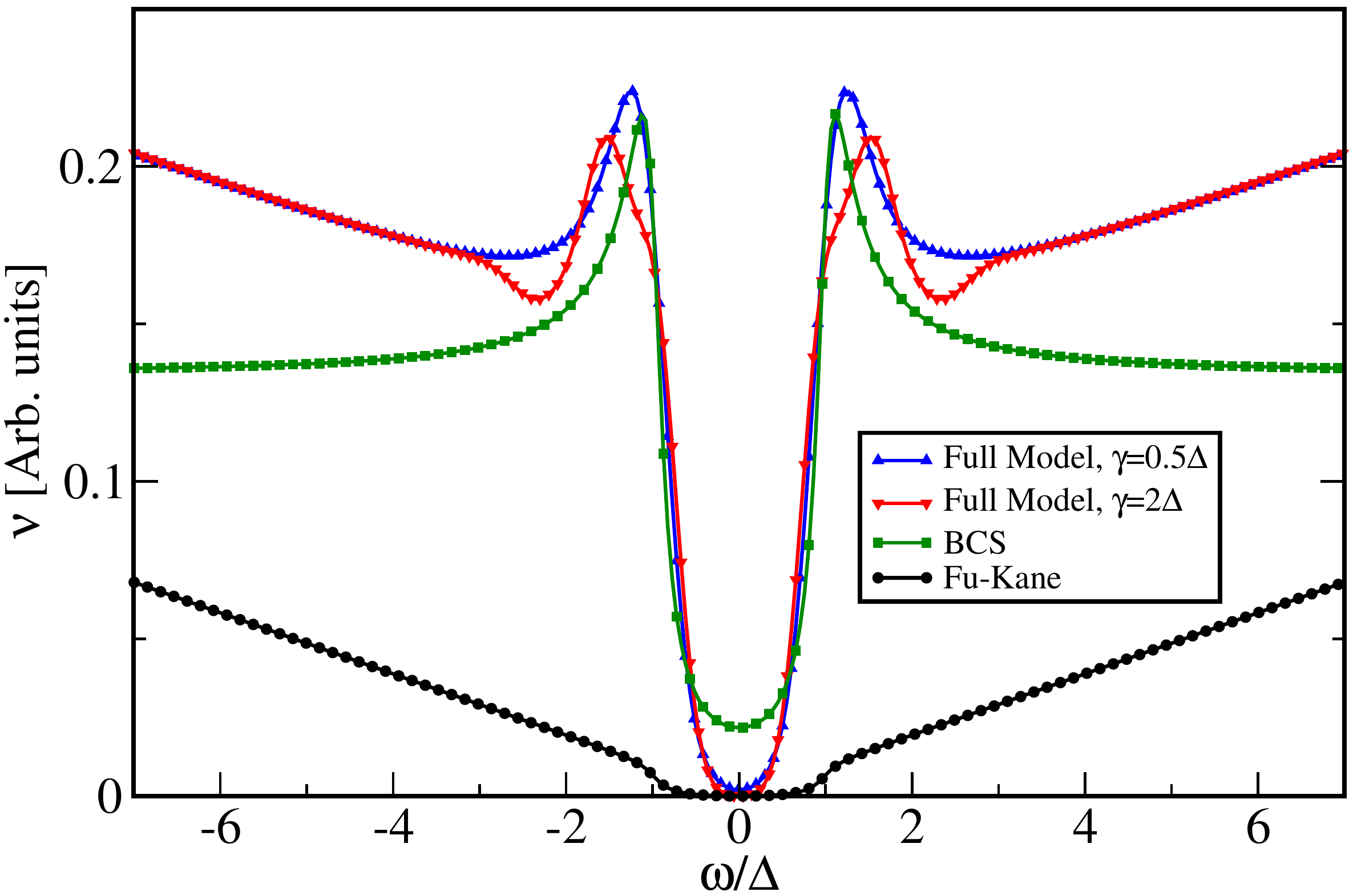}
\caption{A comparison of the density of states for the Fu-Kane model (black circles), the standard BCS model (green squares), and our full model for $\gamma=2\Delta$ (red triangles down), and $\gamma=0.5\Delta$ (blue triangles up). We have used the velocity $v_F=10$ meV{\AA}.}
\label{ref:figure_dos1}
\end{figure}

In figure \ref{ref:figure_dos1} we compare the density of states for our full hybrid model with that of the Fu-Kane model, equation \eqref{fkdos}, and the standard BCS model, equation \eqref{bcsdos}. Although only small differences can be seen between the BCS and hybrid cases inside the gap, the effect becomes very pronounced outside of this where the effect of the Dirac cone becomes very clear. For weak coupling $\gamma$ one finds typical looking coherence peaks in the hybrid case. However as the coupling is increased more structure becomes apparent near the edge of the gap, including a characteristic reduction in the density of states just outside the coherence peaks.

\begin{figure}[t!]
\centering
\includegraphics[width=0.99\columnwidth]{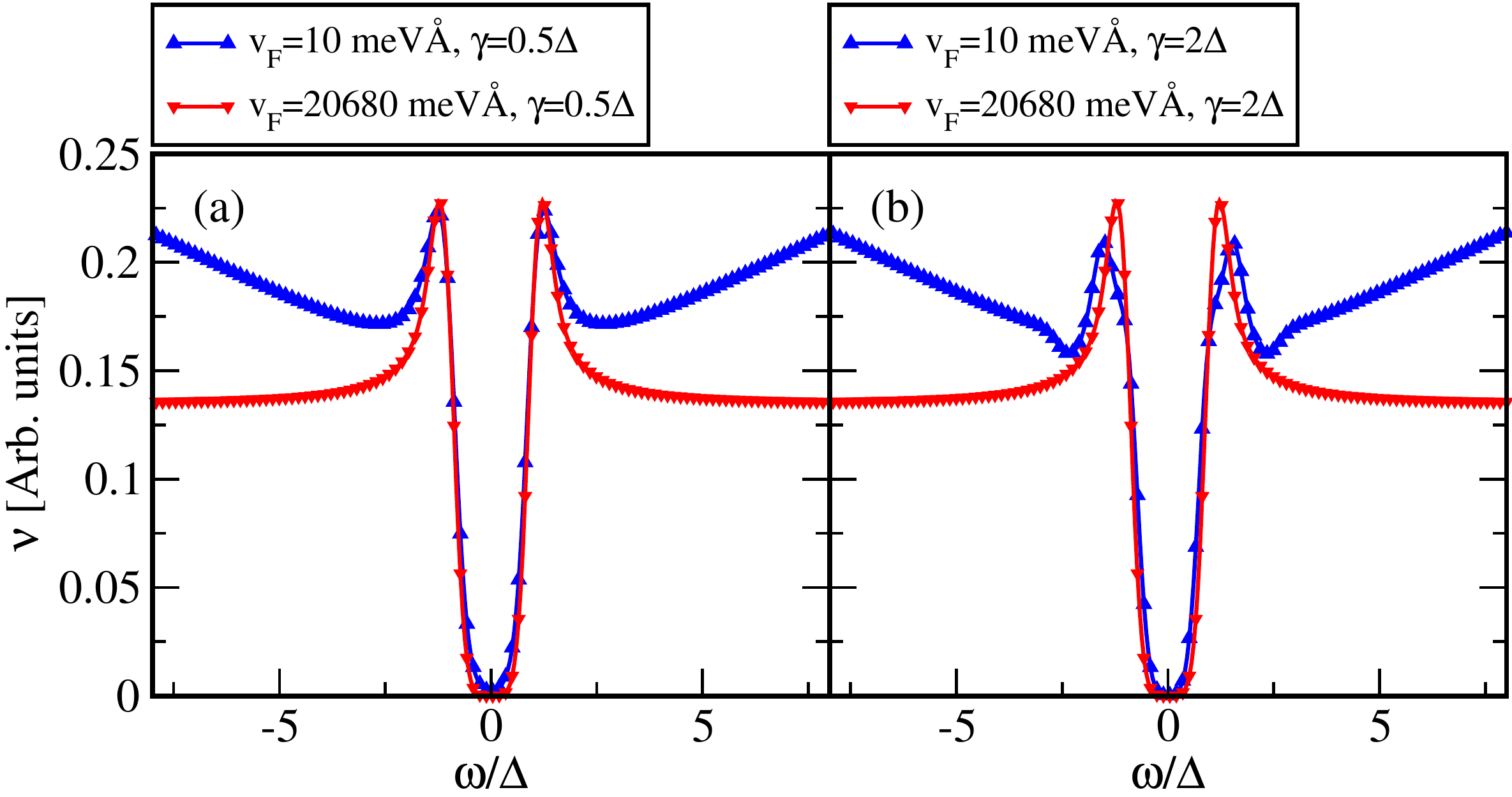}
\caption{A comparison of the density of states for our full model with different coupling strengths $\gamma$ and velocities $v_F$. Panel (a) shows the density of states for $\gamma=0.5\Delta$ with velocities $v_F=10$ meV{\AA} (blue triangles up), and $v_F=20680$ meV{\AA} (red triangles down). Panel (b) shows the density of states for $\gamma=2\Delta$ with velocities $v_F=10$ meV{\AA} (blue triangles up), and $v_F=20680$ meV{\AA} (red triangles down). We note that for the velocity $v_F=20680$ meV{\AA}, as is predicted for Bi$_2$Se$_3$ for example, the density of states of the TPSS is almost flat and hence shows no visible sign for either coupling strength. However the velocity is expected to be much lower in the hybrid structure \citep{Sedlmayr}, and hence we use an exemplary value of $v_F=10$ meV{\AA} in panel (a).}
\label{ref:figure_dos2}
\end{figure}

In figure \ref{ref:figure_dos2} we compare different coupling strengths and Fermi velocities of the TPSS for our hybrid model. We note that for the larger values of the Fermi velocities, $v_F=20680$ meV{\AA}, which corresponds to the values known for Bi$_2$Se$_3$ \citep{Zhang-Zhang}, the density of states of the TPSS is too flat to have any visible effect on the density of states. However, as already mentioned, we expect the Fermi velocity of our model to depend on the thickness of the SC layer, $\ell_{\rm S}$, as
\begin{equation}\label{thick}
    v_F=v_{F0}\beta\frac{\ell_{\rm TI}}{\ell_{\rm TI}+\ell_{\rm S}}\,,
\end{equation}
where $\ell_{\rm TI}$ is the penetration depth of the TPSS into the TI bulk. This is likely to be further reduced by effects of the boundary, which are modelled by the dimensionless parameter $\beta$. If $\ell_{\rm TI}\ll\ell_{\rm S}$ then we have $v_F=v_{F0}\ell_{\rm TI}/(\beta\ell_{\rm S})$ and $v_F\ll v_{F0}$, which is in agreement with experiments \citep{Sedlmayr}. In turn this means that counter-intuitively, the effect is best seen in SC layers of an intermediate thickness. Naturally spin-ARPES measurements would not suffer form this defect, and thinner layers would be preferential in that case.

\begin{figure}[t!]
\centering
\includegraphics[width=0.99\columnwidth]{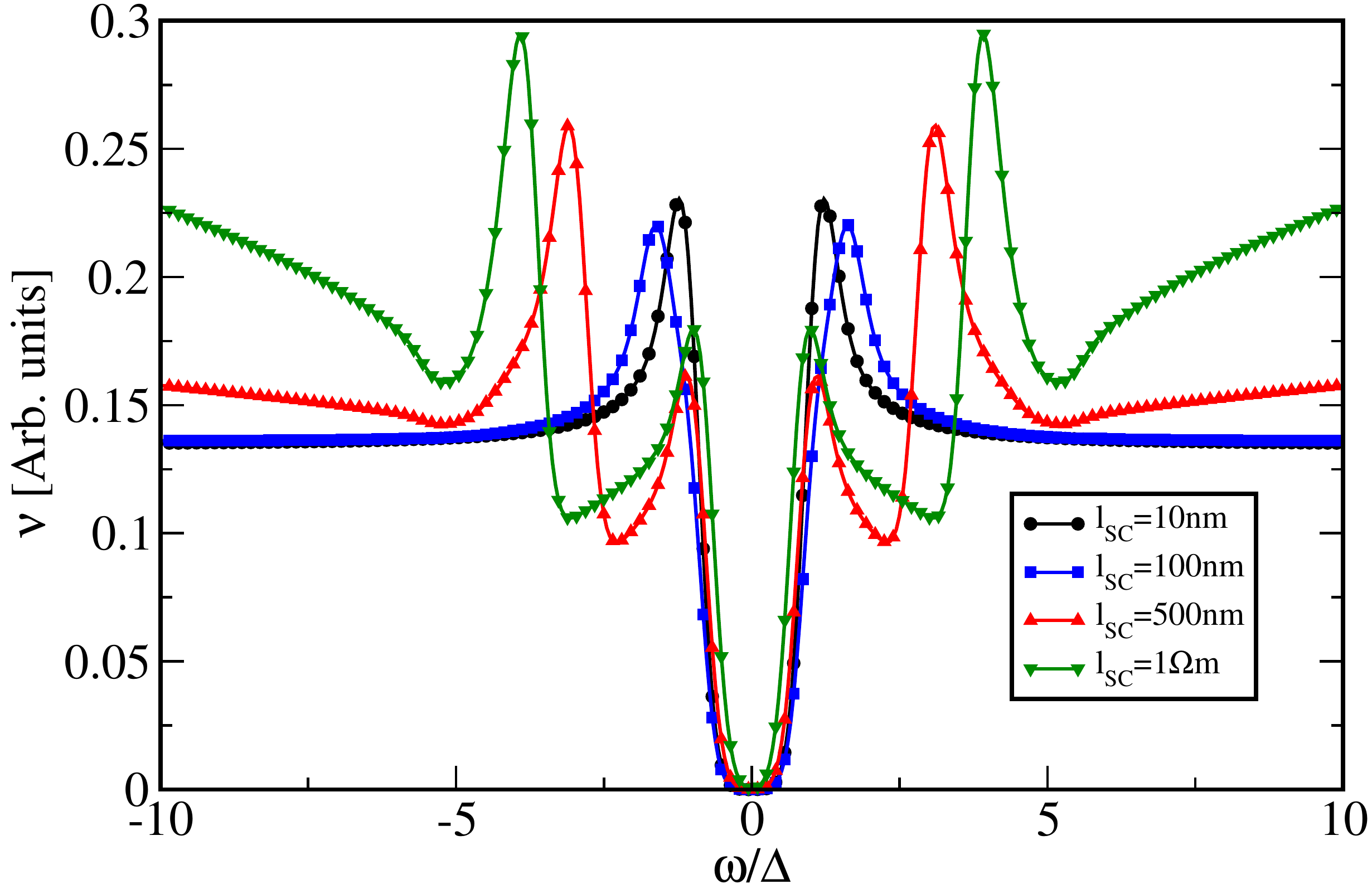}
\caption{A comparison of the density of states for our full hybrid model for $\gamma=5\Delta$ and different thicknesses of the SC layer $\ell_{\rm S}$ where the velocity is given by equation \eqref{thick} with $v_F=20680$ meV{\AA}, $\ell_{\rm TI}=2$nm and $\beta=0.25$. Further features become visible around the coherence peaks of the BCS states as the velocity is reduced.}
\label{ref:figure_dos3}
\end{figure}

At strong coupling, further features appear around the coherence peaks. In figure \ref{ref:figure_dos3} we show results for the large coupling regime $\gamma=5\Delta$ at a series of SC thicknesses, the effects of which are modelled by equation \eqref{thick} with $\ell_{\rm TI}=2$nm and $\beta=0.25$. In a real physical system this effect may be reduced by the reduction in the coupling for thicker SC layers, and may be hard to achieve.

\section{Effective description}\label{sec_eff}

We will derive two effective models for which either the SC or TPSS degree of freedom has been removed. This leads to a modified topological superconductor like model for the TPSS, and to a superconductor with an induced spin-orbit like coupling term. We will use a functional integral representation \cite{Negele-Orland} in the Matsubara formalism. For this purpose it is convenient to first diagonalize the SC Hamiltonian $H_{\rm S}$. Let $\widehat{\Psi}_{\bf{k}}=\mathcal{U}_k\widetilde{\Psi}_{\bf{k}}$ with the rotation matrix
\begin{equation}
 \mathcal{U}_k=\frac{1}{\sqrt{2}}
 \begin{pmatrix}
 0&\frac{\Delta}{\alpha^-_k}&0&-\frac{\Delta}{\alpha^+_k}\\
 \frac{\Delta}{\alpha^-_k}&0&-\frac{\Delta}{\alpha^+_k}&0\\
 0&\frac{\alpha^-_k}{\epsilon_k}&0&\frac{\alpha^+_k}{\epsilon_k}&\\
  \frac{\alpha^-_k}{\epsilon_k}&0&\frac{\alpha^+_k}{\epsilon_k}&0
 \end{pmatrix}\,,
\end{equation}
where $\alpha^\pm_k=\sqrt{\epsilon_k^2\pm\epsilon_k\xi_k}$.  This results in the diagonal Hamiltonian density
\begin{equation}\label{bcs}
\widehat\h_{\rm S}=\epsilon_k{\bm \tau}^z\,.
\end{equation}
In this new basis the coupling term is
\begin{equation}\label{coupling}
\widehat H_{\rm C}=
\int\ud^2{\bf{k}}\widetilde\chi^\dagger_{\bf{k}}\bm{\Gamma}_k\widehat\Psi_{\bf{k}}+\textrm{H.c.}\,,
\quad \bm{\Gamma}_k=\gamma{\bm \tau}^z\mathcal{U}_k.
\end{equation}
We make no changes to the field operators for the TPSS.

In the following $\Psi_{n{\bf{k}}}$ and $\chi_{n{\bf{k}}}$ are Grassmann fields corresponding to the fermionic field operators $\widehat\Psi_{{\bf{k}}}$ and $\widetilde\chi_{{\bf{k}}}$ respectively. The functional integrals we will need to perform are over quadratic actions and are given by
\begin{align}
\left\langle \ldots\right\rangle_{{\rm S}}=&\frac{\int D{\Psi}\ldots\e^{-S_{\rm S}}}{\int D{\Psi}\e^{-S_{\rm S}}}\textrm{, where}\\\nonumber
S_{\rm S}=&-T\sum_n\int\ud^2{\bf{k}}\Psi^\dagger_{n{\bf{k}}}\left[\omega_n+\widehat\h_{\rm S}\right]\Psi_{n{\bf{k}}}\,,
\end{align}
for the SC and
\begin{align}
\left\langle \ldots\right\rangle_{{\rm TI}}=&\frac{\int D{\chi}\ldots\e^{-S_{\rm TI}}}{\int D{\chi}\e^{-S_{\rm Ti}}}\textrm{, where}\\\nonumber
S_{\rm TI}=&-T\sum_n\int\ud^2{\bf{k}}\chi^\dagger_{n{\bf{k}}}\left[\omega_n+\widetilde\h_{\rm TI}\right]\chi_{n{\bf{k}}}\,,
\end{align}
for the TPSS.

\subsection{Dual PE effective model for the TI states}\label{sec_eff_fk}

To find our effective model for the TPSS with the SC states integrated out we must calculate the correction to $\widetilde\h_{\rm TI}$ from the coupling term with the BCS states integrated out. We therefore want to calculate
\begin{equation}
\e^{-S'}=\left\langle \e^{-T\sum_n\int\ud^2{\bf{k}}\left[\chi^\dagger_{n{\bf{k}}}{\bm \Gamma}_k\Psi_{n{\bf{k}}}+\textrm{H.c.}\right]}\right\rangle_{{\rm S}}\,.
\end{equation}
After performing the functional integral, and using standard formula that relates determinant of an operator to its trace, $\ln \mathrm{det} O=\mathrm{tr}\ln O$, we find
\begin{equation}
S'=\frac{1}{2}T\sum_{n}\int\ud^2{\bf{k}}
\left[\chi^\dagger_{n{\bf{k}}}{\bm \Gamma}_k^\dagger {\bm g}_{nk}{\bm \Gamma}_k\chi_{n{\bf{k}}}+\textrm{H.c.}\right]\,,
\end{equation}
where ${\bm g}_{nk}=\langle\Psi_{nk}\Psi^\dagger_{nk}\rangle_{\rm S}$ is the Green's function for the superconducting states. This term is in addition to $\widetilde\h_{\rm TI}$ and we therefore have for the effective Hamiltonian density
\begin{equation}
\h^{\rm eff}_{\rm TI}=\widetilde\h_{\rm TI}+\Delta^{\rm eff}_{nk}{\bm \tau}^x-\mu^{\rm eff}_{nk}{\bm \tau}^z\,,
\end{equation}
where
\begin{equation}
\Delta^{\rm eff}_{nk}=\frac{\gamma^2\Delta}{\omega_n^2+\epsilon_k^2}\,, \quad\textrm{and}\quad
\mu^{\rm eff}_{nk}=\frac{\gamma^2\xi_k}{\omega_n^2+\epsilon_k^2}\,.
\end{equation}
Thus in the static limit we are left with the effective model:
\begin{align}\label{effh}
\h^{\rm eff}=&(v_Fk{\bm\sigma}^z-\mu^{\rm eff}_{k}){\bm \tau}^z+\Delta^{\rm eff}_{k}{\bm \tau}^x\,,\\\nonumber
\Delta^{\rm eff}_{k}=&\frac{\gamma^2\Delta}{\Delta^2+\xi^2_k}\,,\quad \textrm{and}\quad
\mu^{\rm eff}_{k}={\tilde{\mu}}+\frac{\gamma^2\xi_k}{\Delta^2+\xi_k^2}\,.
\end{align}
The effective s-wave pairing near the Fermi level is therefore $\sim\gamma^2/\Delta$. For an appreciable pairing strength in this model the hybridisation must therefore be similar in magnitude to the s-wave gap of the superconductor.

Close to the centre of the Dirac cone the Fu-Kane model \cite{Fu-Kane} is recovered, i.e. with $\Delta^{\rm eff}_{k}\to \Delta$
\begin{equation}
\h_{\rm FK}=(v_Fk{\bm\sigma}^z-{\tilde{\mu}}){\bm \tau}^z+\Delta{\bm \tau}^x\,.
\end{equation}
As discussed earlier this model is known to host MZM at the centre of vortices.

\begin{figure}[t!]
\centering
\includegraphics[width=0.99\columnwidth]{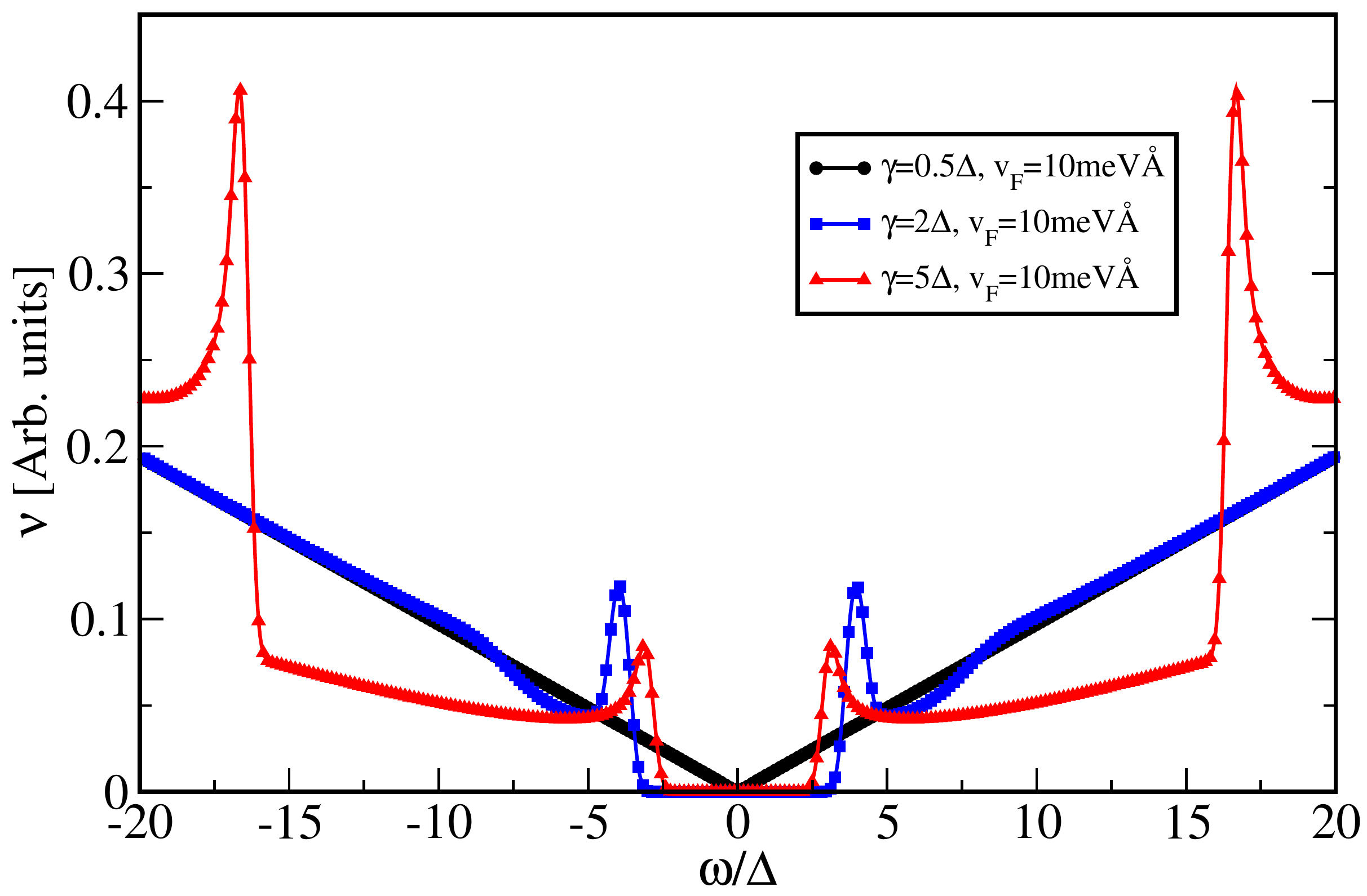}
\caption{A comparison of the density of states for the effective TPSS model equation \eqref{effh}. A gap is induced by coupling to the TPSS, and can induce features at higher energies which are similar to secondary coherence peaks.}
\label{ref:figure_dos4}
\end{figure}

The effective TPSS model \eqref{effh} has eigenenergies
\begin{equation}
    \varepsilon^{ab}_{\rm{TI};k}=\frac{a}{\epsilon_k}\sqrt{v_F^2k^2\Delta^2+(v_Fk\xi_k+b\gamma^2)^2}
\end{equation}
with $a=\pm1$ and $b=\pm1$. As in the case of the full hybrid model, using \eqref{fulldos} we can calculate the corresponding density of states for this model, see figure \ref{ref:figure_dos4}. By increasing the coupling strength top the SC states one finds an induced gap, as expected, with coherence peaks. Additional feature can be seen outside of this region, which for stronger coupling become more pronounced, leading to what has the appearance of a second coherence peak.

\subsection{Dual PE effective model for the S states}\label{sec_eff_bcs}

\begin{figure}[t!]
\centering
\includegraphics[width=0.99\columnwidth]{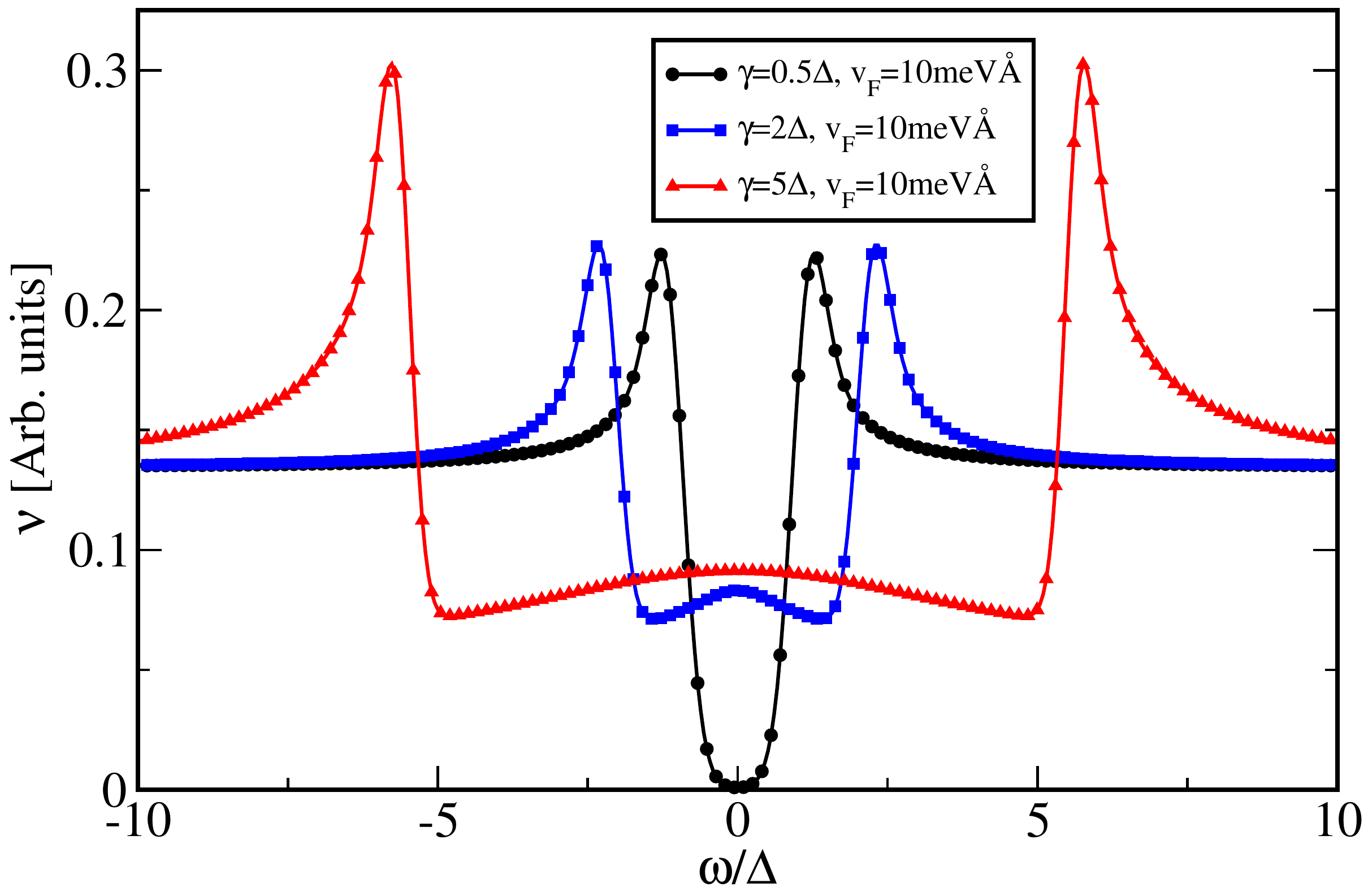}
\caption{A comparison of the density of states for the effective SC model equation \eqref{effhsc}. The coupling to the TPSS widens the superconducting gap and generates mid-gap states.}
\label{ref:figure_dos5}
\end{figure}

Similarly the SC will be modified by the presence of the TPSS. In this case we need to calculate the action $S''$ from
\begin{equation}
\e^{-S''}=\left\langle \e^{-T\sum_n\int\ud^2{\bf{k}}\left[\chi^\dagger_{n{\bf{k}}}{\bm \Gamma}_k\Psi_{n{\bf{k}}}+\textrm{H.c.}\right]}\right\rangle_{{\rm TI}}\,.
\end{equation}
In complete analogy to the preceding calculation, following this standard integral we find the effective Hamiltonian
\begin{equation}\label{effhsc}
\h^{\rm eff}_{\rm S}=\widehat\h_{\rm S}+\left[\Sigma^{\rm eff}_{nk}-\alpha^{\rm eff}_{nk}k{\bm\sigma}^y\right]{\bm\tau}^z\,.
\end{equation}
where
\begin{equation}
 \Sigma^{\rm eff}_{nk}=\frac{\gamma^2\mu(\mu^2+\omega_n^2-v_F^2k^2)}{4\mu^2\omega_n^2+[\mu^2-\omega_n^2-v_F^2k^2]^2}\,,
\end{equation}
 and
 \begin{equation}
 \alpha^{\rm eff}_{nk}=\frac{\gamma^2v_F(\mu^2-\omega_n^2-v_F^2k^2)}{4\mu^2\omega_n^2+[\mu^2-\omega_n^2-v_F^2k^2]^2}\,.
\end{equation}
$\alpha^{\rm eff}_{nk}$ is an effective spin-orbit like term introduced by the TPSS. In the static limit this gives
\begin{align}
 \Sigma^{\rm eff}_{k}=\frac{\gamma^2\mu}{\mu^2-v_F^2k^2}
 \textrm{ and }
 \alpha^{\rm eff}_{k}=\frac{\gamma^2v_F}{\mu^2-v_F^2k^2}\,.
\end{align}
This effective model for the SC has eigenenergies
\begin{equation}
    \varepsilon^{ab}_{\rm{S};k}=a \epsilon_k+a\frac{\gamma^2}{\mu+b v_F k}
\end{equation}
with $a=\pm1$ and $b=\pm1$. As previously we can use \eqref{fulldos} to calculate the density of states, see figure \ref{ref:figure_dos5}. The effect of the TPSS as the coupling strength is increased is to both widen the gap and generate mid-gap states.

\section{Summary}\label{Sec:Summary}

In this communication we have introduced and solved a minimal model which describes the interplay between SC and TPSS in a thin superconducting film on top of a topological insulator. Due to a dual proximity effect TPSS from the TI tunnel into the SC, and conversely Cooper pairs from the SC tunnel into the TI. Our model describes the hybridization between these states. We show that this gives rise to low energy bands with the appropriate SC gap and spin momentum locking to potentially give rise to a topological superconductor. Second we calculate the density of states of this full hybrid model, and demonstrate that it has distinctive features which could be measured in STM experiments. We benchmark this density of states to the standard BCS density of states and that for the minimal Fu-Kane model of $s$-wave proximitized TPSS. The hybridization leads to a modified gap structure and distinctive slopes outside of the gapped region which originate from the TPSS. We also find that the coherence peaks become modified.

We then consider two effective models which describe the effect of the SC states on the TPSS, and the TPSS on the SC states. We find that for the TPSS a modified Fu-Kane like model is formed, with a momentum dependent gap. In the limit $k\to0$ this tends toward the standard Fu-Kane result. The effect of the TPSS on the SC is rather to introduce a complicated spin-orbit coupling term, due to the spin-momentum locking of the TPSS. For both models we derive an effective Hamiltonian, the resulting band structure, and the density of states. We demonstrate that the effective TPSS model density of states has a proximity induced gap and shows coherence peaks which are absent from the minimal Fu-Kane model. Further features can also appear in the case of strong hybridization. For the SC, the TPSS tend to widen the gap and coherence peaks, and induce mid-gap states. All of these features would be visible in STM experiments.

There are several possible extensions of this work which would be interesting to consider. One limitation of the full hybrid model which is introduced here, is that it is strictly two dimensional. A future extension of this work would be to solve the full three dimensional strongly-coupled system self-consistently, for instance by generalizing the earlier analysis of Ref. \cite{Lababidi}, and investigate more thoroughly the role played by the thickness of the SC and TI, as well as more general coupling terns at the interface. Similarly, how a more realistic band structure of the TI would affect the results and how disorder in the materials would change the density of states and proximity effects is also of interest. If the final goal is to demonstrate the possible existence of MZMs, then the effects of the magnetic field which causes the vortices would also need to be taken into account.

\section*{Acknowledgements}

We gratefully acknowledge discussions with Paulo Sessi, Stuart Tessmer, Eric Goodwin, and Dale van Harlingen.

Support for this work at the University of Wisconsin-Madison was provided by the National Science Foundation, Quantum Leap Challenge Institute for Hybrid Quantum Architectures and Networks, NSF Grant No. 2016136. 

In M. Curie-Sk{\l}odowska University N.S. was supported by the National Science Centre (NCN, Poland) under Grant No.~UMO-2019/35/B/ST3/03625.

\printcredits

\bibliographystyle{model1-num-names}

\begin{thebibliography}{99}



\bibitem{Kitaev}
A. Yu. Kitaev,
\textit{Fault-tolerant quantum computation by anyons},
Annals of Physics \textbf{303} (2003) 2.

\bibitem{Nayak}
Chetan Nayak, Steven H.  Simon, Ady Stern, Michael Freedman, and Sankar Das Sarma,
\textit{Non-Abelian anyons and topological quantum computation}, Rev. Mod. Phys. \textbf{80} (2008) 1083.

\bibitem{Golubov}
A. A. Golubov, M. Yu. Kupriyanov, and E. Il'ichev,
\textit{The current-phase relation in Josephson junctions},
Rev. Mod. Phys. \textbf{76} (2004) 411.

\bibitem{Beenakker}
C. W. J. Beenakker,
\textit{Random-matrix theory of Majorana fermions and topological superconductors},
Rev. Mod. Phys. \textbf{87} (2015) 1037.

\bibitem{Lutchyn}
R. M. Lutchyn, J. D. Sau, and S. Das Sarma,
\textit{Majorana Fermions and a Topological Phase Transition in Semiconductor-Superconductor Heterostructures},
Phys. Rev. Lett. \textbf{105} (2010) 077001.

\bibitem{Oreg}
Y. Oreg, G. Refael, and F. von Oppen,
\textit{Helical Liquids and Majorana Bound States in Quantum Wires},
Phys. Rev. Lett. \textbf{105} (2010) 177002.

\bibitem{Pientka}
Falko Pientka, Anna Keselman, Erez Berg, Amir Yacoby, Ady Stern, and Bertrand I. Halperin,
\textit{Topological Superconductivity in a Planar Josephson Junction},
Phys. Rev. X \textbf{7} (2017) 021032.

\bibitem{Ivanov}
D. A. Ivanov, \textit{Non-Abelian Statistics of Half-Quantum Vortices in
$p$-Wave Superconductors}, Phys. Rev. Lett. \textbf{86} (2001) 268.

\bibitem{Fu-Kane}
Liang Fu and C. L. Kane,
\textit{Superconducting Proximity Effect and Majorana Fermions at the Surface of a Topological Insulator},
Phys. Rev. Lett. \textbf{100} (2008) 096407.

\bibitem{Hughes}
Xiao-Liang Qi, Taylor L. Hughes, S. Raghu, and Shou-Cheng Zhang, \textit{Time-Reversal-Invariant Topological Superconductors and Superfluids in Two and Three Dimensions}, Phys. Rev. Lett. \textbf{102} (2009) 187001.

\bibitem{Ioselevich}
P. A. Ioselevich, P. M. Ostrovsky, and M. V. Feigel'man,
\textit{Majorana state on the surface of a disordered 3D topological insulator},
Phys. Rev. B \textbf{86} (2012) 035441.

\bibitem{Qi-Zhang}
Xiao-Liang Qi and Shou-Cheng Zhang,
\textit{Topological insulators and superconductors},
Rev. Mod. Phys. \textbf{83} (2011) 1057.

\bibitem{Vishveshwara}
Eytan Grosfeld, Babak Seradjeh, and Smitha Vishveshwara,
\textit{Proposed Aharonov-Casher interference measurement of non-Abelian vortices in chiral $p$-wave superconductors},
Phys. Rev. B \textbf{83} (2001) 104513.

\bibitem{Stern}
Eytan Grosfeld and Ady Stern,
\textit{Observing Majorana bound states of Josephson vortices in topological superconductors},
Proc. Natl. Acad. Sci. U.S.A. \textbf{10} (2011)11810.

\bibitem{Hegde}
Suraj S. Hegde, Yuxuan Wang, Erik Huemiller, Guang Yue, D. J. Van Harlingen, and Smitha Vishveshwara,
\textit{A topological Josephson junction platform for creating, manipulating, and braiding Majorana bound states},
preprint arXiv:1907.02935 [cond-mat.mes-hall, cond-mat.supr-con].


\bibitem{Mourik}
V. Mourik, K. Zuo, S. M. Frolov, S. R. Plissard, E. P. A. M. Bakkers, and L. P. Kouwenhoven,
\textit{Signatures of Majorana Fermions in Hybrid Superconductor-Semiconductor Nanowire Devices},
Science  \textbf{336} (2012) 1003.

\bibitem{Das}
Anindya Das, Yuval Ronen, Yonatan Most, Yuval Oreg, Moty Heiblum, and Hadas Shtrikman,
\textit{Zero-bias peaks and splitting in an Al-InAs nanowire topological superconductor as a signature of Majorana fermions},
Nat. Phys. \textbf{8} (2012) 887.

\bibitem{Deng}
M. T. Deng, C. L. Yu, G. Y. Huang, M. Larsson, P. Caroff, and H. Q. Xu,
\textit{Anomalous Zero-Bias Conductance Peak in a Nb–InSb Nanowire–Nb Hybrid Device},
Nano Lett. \textbf{12} (2012) 6414.

\bibitem{Albrecht}
S. M. Albrecht, A. P. Higginbotham, M. Madsen, F. Kuemmeth, T. S. Jespersen, J. Nyg\aa rd, P. Krogstrup, and C. M. Marcus,
\textit{Exponential protection of zero modes in Majorana islands},
Nature \textbf{531} (2016) 206.

\bibitem{Suominen}
H. J. Suominen, M. Kjaergaard, A. R. Hamilton, J. Shabani, C. J. Palmstr\o m, C. M. Marcus, and F. Nichele,
\textit{Zero-Energy Modes from Coalescing Andreev States in a Two-Dimensional Semiconductor-Superconductor Hybrid Platform},
Phys. Rev. Lett. \textbf{119} (2017) 176805.

\bibitem{Gul}
\"Onder G\"ul, Hao Zhang, Jouri D. S. Bommer, Michiel W. A. de Moor, Diana Car, S\'ebastien R. Plissard, Erik P. A. M. Bakkers, Attila Geresdi, Kenji Watanabe, Takashi Taniguchi and Leo P. Kouwenhoven, \textit{Ballistic Majorana nanowire devices},
Nat. Nano. \textbf{13} (2018) 192.

\bibitem{Zhang}
Hao Zhang, Chun-Xiao Liu, Sasa Gazibegovic, Di Xu, John A Logan, Guanzhong Wang, Nick van Loo, Jouri D S Bommer, Michiel W. A. de Moor, Diana Car, Roy L. M. Op het Veld, Petrus J. van Veldhoven, Sebastian Koelling, Marcel A Verheijen, Mihir Pendharkar, Daniel J Pennachio, Borzoyeh Shojaei, Joon Sue Lee, Chris J Palmstr\o m, Erik P A M Bakkers, S. Das Sarma, and Leo P Kouwenhoven, \textit{Quantized Majorana conductance}, Nature \textbf{556} (2018) 74.

\bibitem{Larsen}
T. W. Larsen, K. D. Petersson, F. Kuemmeth, T. S.  Jespersen, P. Krogstrup, J. Nyg\aa rd, and C. M. Marcus,
\textit{Semiconductor-Nanowire-Based Superconducting Qubit},
Phys. Rev. Lett. \textbf{115} (2015) 127001.

\bibitem{Casparis}
L. Casparis, T. W. Larsen, M. S. Olsen, F. Kuemmeth, P. Krogstrup, J. Nyg\aa rd, K. D. Petersson, and C. M. Marcus,
\textit{Gatemon Benchmarking and Two-Qubit Operations},
Phys. Rev. Lett. \textbf{116} (2016) 150505.

\bibitem{Vavilov}
Zhenyi Qi, Hong-Yi Xie, Javad Shabani, Vladimir E. Manucharyan, Alex Levchenko, and Maxim G. Vavilov,
\textit{Controlled-Z gate for transmon qubits coupled by semiconductor junctions},
Phys. Rev. B 97 (2018) 134518.

\bibitem{Chen}
J. Chen, B. D. Woods, P. Yu, M. Hocevar, D. Car, S. R. Plissard, E. P. A. M. Bakkers, T. D. Stanescu, and S. M. Frolov,
\textit{Ubiquitous Non-Majorana Zero-Bias Conductance Peaks in Nanowire Devices},
Phys. Rev. Lett. \textbf{123}  (2019) 107703.

\bibitem{Yu}
P. Yu, J. Chen, M. Gomanko, G. Badawy, E. P. A. M. Bakkers, K. Zuo, V. Mourik, and S. M. Frolov,
\textit{Non-Majorana states yield nearly quantized conductance in superconductor-semiconductor nanowire devices},
preprint arXiv:2004.08583 [cond-mat.mes-hall, cond-mat.supr-con].

\bibitem{Aguado1}
Elsa Prada, Pablo San-Jose, Michiel W. A. de Moor, Attila Geresdi, Eduardo J. H. Lee, Jelena Klinovaja, Daniel Loss, Jesper Nygard, Ram\'on Aguado and Leo P. Kouwenhoven, \textit{From Andreev to Majorana bound states in hybrid superconductor-semiconductor nanowires}, Nature Reviews Physics \textbf{2}, (2020) 575

\bibitem{Aguado2}
J. Avila, F. Pe{\~{n}}aranda, E. Prada, P. San-Jose and R. Aguado, \textit{Non-hermitian topology as a unifying framework for the Andreev versus Majorana states controversy}, Commun Phys, \textbf{2}, (2019) 133

\bibitem{Pan}
Haining Pan, William S. Cole, Jay D. Sau, and S. Das Sarma, \textit{Generic quantized zero-bias conductance peaks in superconductor-semiconductor hybrid structures}, Phys.~Rev.~B \textbf{101} (2020) 024506

\bibitem{Hendrickx}
N. Hendrickx, D. Franke, A. Sammak, M. Kouwenhoven, D. Sabbagh, L. Yeoh, R. Li, M. Tagliaferri, M.  Virgilio, G. Capellini, G. Scappucci, and M. Veldhorst, \textit{Gate-controlled quantum dots and superconductivity in planar germanium}, Nat. Comm.  \textbf{9} (2018) 2835.

\bibitem{Vigneau}
Florian Vigneau, Raisei Mizokuchi, Dante Colao Zanuz, Xuhai Huang, Susheng Tan, Romain Maurand, Sergey Frolov, Amir Sammak, Giordano Scappucci, Francois Lefloch, and Silvano De Franceschi, \textit{Germanium Quantum-Well Josephson Field-Effect Transistors and Interferometers}, Nano Lett. \textbf{2} (2019) 1023.


\bibitem{Brinkman}
M. Veldhorst, M. Snelder, M. Hoek, T. Gang, V. K. Guduru, X. L. Wang, U. Zeitler, W. G Van Der Wiel, A. A. Golubov, H. Hilgenkamp, and A. Brinkman, \textit{Josephson supercurrent through a topological insulator surface state}, Nature Materials \textbf{11} (2012) 417.

\bibitem{Goldhaber-Gordon}
J. R. Williams, A. J. Bestwick, P. Gallagher, Seung Sae Hong, Y. Cui, Andrew S. Bleich, J. G. Analytis, I. R. Fisher, and D. Goldhaber-Gordon, \textit{Unconventional Josephson Effect in Hybrid Superconductor-Topological Insulator Devices}, Phys. Rev. Lett. \textbf{109} (2012) 056803.

\bibitem{Molenkamp}
Luis Maier, Jeroen B. Oostinga, Daniel Knott, Christoph Br\"une, Pauli Virtanen, Grigory Tkachov, Ewelina M. Hankiewicz, Charles Gould, Hartmut Buhmann, and Laurens W. Molenkamp, \textit{Induced Superconductivity in the Three-Dimensional Topological Insulator HgTe},
Phys. Rev. Lett. \textbf{109} (2012) 186806.

\bibitem{LiLu}
Fanming Qu, Fan Yang, Jie Shen, Yue Ding, Jun Chen, Zhongqing Ji, Guangtong Liu, Jie Fan, Xiunian Jing, Changli Yang, and Li Lu, \textit{Strong Superconducting Proximity Effect in Pb-Bi$_2$Te$_3$ Hybrid Structures},
Scientific Reports \textbf{2} (2012) 339.

\bibitem{Mason}
S. Cho, B. Dellabetta, A. Yang, J. Schneeloch, Z. Xu, T. Valla, G. Gu. M. J. Gilbert, and N. Mason, \textit{Symmetry protected Josephson supercurrents in three-dimensional topological insulators}, Nat. Comm. \textbf{4}, (2013) 1689.

\bibitem{Oostinga}
Jeroen B. Oostinga, Luis Maier, Peter Sch\"uffelgen, Daniel Knott, Christopher Ames, Christoph Br\"une, Grigory Tkachov, HartmutBuhmann, and Laurens W. Molenkamp, \textit{Josephson Supercurrent through the Topological Surface States of Strained Bulk HgTe},
Phys. Rev. X \textbf{3} (2013) 021007.

\bibitem{Kurter-1}
C. Kurter, A. D. K. Finck, P. Ghaemi, Y. S. Hor, and D. J. Van Harlingen, \textit{Dynamical gate-tunable supercurrents in topological Josephson junctions}, Phys. Rev. B \textbf{90} (2014) 014501.

\bibitem{Kurter-2}
C. Kurter, A. D. K. Finck, Y. S. Hor, and D. J. Van Harlingen, \textit{Evidence for an anomalous current–phase relation in topological insulator Josephson junctions}, Nat. Comm. \textbf{6} (2015) 7130.

\bibitem{Sochnikov}
Ilya Sochnikov, Luis Maier, Christopher A. Watson, John R. Kirtley, Charles Gould, Grigory Tkachov, Ewelina M. Hankiewicz, Christoph Br\"une, Hartmut Buhmann, Laurens W. Molenkamp, and Kathryn A. Moler, \textit{Nonsinusoidal Current-Phase Relationship in Josephson Junctions from the 3D Topological Insulator HgTe}, Phys. Rev. Lett. \textbf{114} (2015) 066801.

\bibitem{Stehno}
M. P. Stehno, V. Orlyanchik, C. D. Nugroho, P. Ghaemi, M. Brahlek, N. Koirala, S. Oh, and D. J. Van Harlingen,
\textit{Signature of a topological phase transition in the Josephson supercurrent through a topological insulator},
Phys. Rev. B 93 (2016) 035307.

\bibitem{Williams}
R. A. Snyder, C. J. Trimble, C. C. Rong, P. A. Folkes, P. J. Taylor, and J. R. Williams,
\textit{Weak-link Josephson Junctions Made from Topological Crystalline Insulators},
Phys. Rev. Lett. \textbf{121} (2018) 097701.

\bibitem{Rokhinson}
Morteza Kayyalha, Mehdi Kargarian, Aleksandr Kazakov, Ireneusz Miotkowski, Victor M. Galitski, Victor M. Yakovenko, Leonid P. Rokhinson, and Yong P. Chen, \textit{Anomalous Low-Temperature Enhancement of Supercurrent in Topological-Insulator Nanoribbon Josephson Junctions: Evidence for Low-Energy Andreev Bound States}, Phys. Rev. Lett. \textbf{122} (2019) 047003

\bibitem{Ando-TSC}
Satoshi Sasaki, M. Kriener, Kouji Segawa, Keiji Yada, Yukio Tanaka, Masatoshi Sato, and Yoichi Ando, \textit{Topological Superconductivity in Cu$_x$Bi$_2$Se$_3$}, Phys. Rev. Lett. \textbf{107} (2011) 217001.

\bibitem{Finck-TSC}
Cihan Kurter, Aaron Finck, Erik Huemiller, Julia Medvedeva, Adam Weis, Juan Atkinson, Yunsheng Qiu, Li Shen, Seng Huat Lee, Thomas Vojta, Pouyan Ghaemi, Yew San Hor, and Dale Van Harlingen, \textit{Andreev Reflection Spectroscopy of Topological Superconductor Candidate Nb$_x$Bi$_2$Se$_3$}, Nano Lett. \textbf{19} (2019) 38.

\bibitem{S-TI-Coexistence-1}
Mei-Xiao Wang, Canhua Liu, Jin-Peng Xu, Fang Yang, Lin Miao, Meng-Yu Yao, C. L. Gao, Chenyi Shen, Xucun Ma, X. Chen, Zhu-An Xu, Ying Liu, Shou-Cheng Zhang, Dong Qian, Jin-Feng Jia, and Qi-Kun Xue, \textit{The Coexistence of Superconductivity and Topological Order in the Bi$_2$Se$_3$ Thin Films}, Science \textbf{336} (2012) 52.

\bibitem{S-TI-Coexistence-2}
Jin-Peng Xu, Canhua Liu, Mei-Xiao Wang, Jianfeng Ge, Zhi-Long Liu, Xiaojun Yang, Yan Chen, Ying Liu, Zhu-An Xu, Chun-Lei Gao, Dong Qian, Fu-Chun Zhang, and Jin-Feng Jia, \textit{Artificial Topological Superconductor by the Proximity Effect},
Phys. Rev. Lett. \textbf{112} (2014) 217001.

\bibitem{S-TI-Coexistence-3}
Hao Yang, Yao-Yi Li, Teng-Teng Liu, Huan-Yi Xue, Dan‐Dan Guan, Shi-Yong Wang, Hao Zheng, Can-Hua Liu, Liang Fu, and Jin-Feng Jia,
\textit{Superconductivity of Topological Surface States and Strong Proximity Effect in Sn$_{1-x}$Pb$_x$Te--Pb Heterostructures},
Advanced Materials \textbf{31} (2019) 1905582.

\bibitem{Burch}
Parisa Zareapour, Alex Hayat, Shu Yang F. Zhao, Michael Kreshchuk, Achint. Jain, Daniel C.  Kwok, Nara Lee, Sang-Wook Cheong, Zhijun Xu, Alina Yang, G. D. Gu, Shuang Jia, Robert J. Cava, and Kenneth S. Burch,  \textit{Proximity-induced high-temperature superconductivity in the topological insulators Bi$_2$Se$_3$ and Bi$_2$Te$_3$}, Nature Comm. \textbf{3} (2012) 1056.

\bibitem{Finck-1}
A. D. K. Finck, C. Kurter, Y. S. Hor, and D. J. Van Harlingen, \textit{Phase Coherence and Andreev Reflection in Topological Insulator Devices},
Phys. Rev. X \textbf{4} (2014) 041022

\bibitem{Tikhonov}
E. S. Tikhonov, D. V. Shovkun, M. Snelder, M. P. Stehno, Y. Huang, M. S. Golden, A. A. Golubov, A. Brinkman, and V. S. Khrapai, \textit{Andreev Reflection in an s-Type Superconductor Proximized 3D Topological Insulator},
Phys. Rev. Lett. \textbf{117} (2016) 147001.

\bibitem{Finck-2}
A. D. K. Finck, C. Kurter, E. D. Huemiller, Y. S. Hor, and D. J. Van Harlingen, \textit{Robust Fabry-Perot interference in dual-gated Bi$_2$Se$_3$ devices}, Appl. Phys. Lett. \textbf{108} (2016) 203101

\bibitem{Banerjee}
Abhishek Banerjee, Ananthesh Sundaresh, Rajamanickam Ganesan, and P. S. Anil Kumar, \textit{Signatures of Topological Superconductivity in Bulk-Insulating Topological Insulator BiSbTe$_{1.25}$Se$_{1.75}$ in Proximity with Superconducting NbSe$_2$}, ACS Nano \textbf{12} (2018) 12665.

\bibitem{MFs-Vortex-BiTeNbSe}
Jin-Peng Xu, Mei-Xiao Wang, Zhi Long Liu, Jian-Feng Ge, Xiaojun Yang, Canhua Liu, Zhu An Xu, Dandan Guan, Chun Lei Gao, Dong Qian, Ying Liu, Qiang-Hua Wang, Fu-Chun Zhang, Qi-Kun Xue, and Jin-Feng Jia, \textit{Experimental Detection of a Majorana Mode in the core of a Magnetic Vortex inside a Topological Insulator-Superconductor Bi$_2$Te$_3$/NbSe$_2$ Heterostructure},
Phys. Rev. Lett. \textbf{114} (2015) 017001.

\bibitem{MFs-Vortex-FeTeSe}
Dongfei Wang, Lingyuan Kong, Peng Fan, Hui Chen, Shiyu Zhu, Wenyao Liu, Lu Cao, Yujie Sun, Shixuan Du, John Schneeloch, Ruidan Zhong, Genda Gu, Liang Fu, Hong Ding, and Hong Jun Gao, \textit{Evidence for Majorana bound states in an iron-based superconductor}, Science \textbf{362} (2018) 333.

\bibitem{MFs-Vortex-LiFe}
Qin Liu, Chen Chen, Tong Zhang, Rui Peng, Ya-Jun Yan, Chen-Hao-Ping Wen, Xia Lou, Yu-Long Huang, Jin-Peng Tian, Xiao-Li Dong, Guang-Wei Wang, Wei-Cheng Bao, Qiang-Hua Wang, Zhi-Ping Yin, Zhong-Xian Zhao, and Dong-Lai Feng, \textit{Robust and Clean Majorana Zero Mode in the Vortex Core of High-Temperature Superconductor (Li$_{0.84}$Fe$_{0.16}$)OHFeSe}, Phys. Rev. X \textbf{8} (2018) 041056.
\bibitem{MFs-Vortex-FeSeTe}
T. Machida, Y. Sun, S. Pyon, S. Takeda, Y. Kohsaka, T. Hanaguri, T. Sasagawa, and T. Tamegai, \textit{Zero-energy vortex bound state in the superconducting topological surface state of Fe(Se,Te)}, Nature Materials \textbf{18} (2019) 811.


\bibitem{Kitaev-SC-mirror-Qubit}
Alexei Kitaev, \textit{Protected qubit based on a superconducting current mirror},  arXiv:cond-mat/0609441  [cond-mat.mes-hall, cond-mat.supr-con, quant-ph].

\bibitem{Preskill}
Peter Brooks, Alexei Kitaev, and John Preskill, \textit{Protected gates for superconducting qubits}, Phys. Rev. A \textbf{87} (2013) 052306.

\bibitem{Koch}
D. K. Weiss, Andy C. Y. Li, D. G. Ferguson, and Jens Koch, \textit{Spectrum and coherence properties of the current-mirror qubit}, Phys. Rev. B \textbf{100} (2019) 224507.

\bibitem{Doucot-1}
B. Dou\c cot, M. V. Feigel'man, and L. B. Ioffe, \textit{Topological Order in the Insulating Josephson Junction Array}, Phys. Rev. Lett. \textbf{90} (2003) 107003.

\bibitem{Doucot-2}
B. Dou\c cot, M. V. Feigel'man, L. B. Ioffe, and A. S. Ioselevich
Phys. Rev. B \textbf{71} (2005) 024505.

\bibitem{Schuster}
Andras Gyenis, Pranav S. Mundada, Agustin Di Paolo, Thomas M. Hazard, Xinyuan You, David I. Schuster, Jens Koch, Alexandre Blais, and Andrew A. Houck, \textit{Experimental realization of an intrinsically error-protected superconducting qubit}, preprint arXiv:1910.07542 [quant-ph, cond-mat.mes-hall].

\bibitem{Gershenson-CPQ-NP}
Sergey Gladchenko, David Olaya, Eva Dupont-Ferrier, Benoit Douçot, Lev B. Ioffe, and  Michael E. Gershenson, \textit{Superconducting nanocircuits for topologically protected qubits}, Nature Physics \textbf{5} (2009) 48.

\bibitem{Gershenson-CPQ-PRL}
Matthew T. Bell, Joshua Paramanandam, Lev B. Ioffe, and Michael E. Gershenson, \textit{Protected Josephson Rhombus Chains}, Phys. Rev. Lett. \textbf{112} (2014) 167001.

\bibitem{Marcus-CPQ}
T. W. Larsen, M. E. Gershenson, L. Casparis, A. Kringh\o j, N. J. Pearson, R. P. G. McNeil, F. Kuemmeth, P. Krogstrup, K. D. Petersson, and C. M. Marcus, \textit{Parity-Protected Superconductor-Semiconductor Qubit}, Phys. Rev. Lett. \textbf{125} (2020) 056801.

\bibitem{Gershenson-PPQ}
M. T. Bell, W. Zhang, L. B. Ioffe, and M. E. Gershenson,\textit{Spectroscopic Evidence of the Aharonov-Casher Effect in a Cooper Pair Box}, Phys. Rev. Lett. \textbf{116} (2016) 107002.

\bibitem{Bell-Bifluxon}
Konstantin Kalashnikov, Wen Ting Hsieh, Wenyuan Zhang, Wen-Sen Lu, Plamen Kamenov, Agustin Di Paolo, Alexandre Blais, Michael E. Gershenson, and Matthew Bell, \textit{Bifluxon: Fluxon-Parity-Protected Superconducting Qubit}, PRX Quantum 1 (2020) 010307.


\bibitem{Nazarov}
Roman-Pascal Riwar, Manuel Houzet, Julia S. Meyer, and Yuli V. Nazarov, \textit{Multi-terminal Josephson junctions as topological matter}, Nature Comm. \textbf{7}, 11167 (2016).

\bibitem{Meyer}
Julia S. Meyer and Manuel Houzet, \textit{Nontrivial Chern Numbers in Three-Terminal Josephson Junctions},Phys. Rev. Lett. 119 (2017) 136807.

\bibitem{Xie-1}
Hong-Yi Xie, Maxim G. Vavilov, and Alex Levchenko, \textit{Topological Andreev bands in three-terminal Josephson junctions}, Phys. Rev. B 96 (2017) 161406(R) (2017).

\bibitem{Eriksson}
Erik Eriksson, Roman-Pascal Riwar, Manuel Houzet, Julia S. Meyer, and Yuli V. Nazarov, \textit{Topological transconductance quantization in a four-terminal Josephson junction}, Phys. Rev. B 95 (2017) 075417.

\bibitem{Xie-2}
Hong-Yi Xie, Maxim G. Vavilov, and Alex Levchenko, \textit{Weyl nodes in Andreev spectra of multiterminal Josephson junctions: Chern numbers, conductances, and supercurrents}, Phys. Rev. B 97 (2018) 035443.

\bibitem{Leone}
R. Leone, L. P. L\'evy, and P. Lafarge,
\textit{Cooper-Pair Pump as a Quantized Current Source}, Phys. Rev. Lett. \textbf{100} (2008) 117001.
\bibitem{Houzet}
Manuel Houzet and Julia S. Meyer, \textit{Majorana-Weyl crossings in topological multiterminal junctions}, Phys. Rev. B \textbf{100} (2019) 014521.

\bibitem{Xie-3}
Hong-Yi Xie and Alex Levchenko, \textit{Topological supercurrents interaction and fluctuations in the multiterminal Josephson effect}, Phys. Rev. B \textbf{99} (2019) 094519.

\bibitem{Finkelstein}
Anne W. Draelos, Ming-Tso Wei, Andrew Seredinski, Hengming Li, Yash Mehta, Kenji Watanabe, Takashi Taniguchi, Ivan V. Borzenets, Francois Amet, and Gleb Finkelstein, \textit{Supercurrent Flow in Multiterminal Graphene Josephson Junctions}, Nano Lett. \textbf{2} (2019)1039.

\bibitem{Pribiag}
Gino V. Graziano, Joon Sue Lee, Mihir Pendharkar, Chris Palmstr\o m, and Vlad S. Pribiag, \textit{Transport Studies in a Gate-Tunable Three-Terminal Josephson Junction}, Phys. Rev. B \textbf{101} (2020) 054510.

\bibitem{Manucharyan}
Natalia Pankratova, Hanho Lee, Roman Kuzmin, Maxim Vavilov, Kaushini Wickramasinghe, William Mayer, Joseph Yuan, Javad Shabani, and Vladimir E. Manucharyan, \textit{The multiterminal Josephson effect}, Phys. Rev. X \textbf{10} (2020) 031051.



\bibitem{Wu-Zhang}
Guangfen Wu, Hua Chen, Yan Sun, Xiaoguang Li, Ping Cui, Cesare Franchini, Jinlan Wang, Xing-Qiu Chen and Zhenyu Zhang, \textit{Tuning the vertical location of helical surface states in topological insulator heterostructures via dual-proximity effects}, Scientific Reports \textbf{3} (2013) 1233.

\bibitem{Dayton}
Ian M. Dayton, Nicholas Sedlmayr, Victor Ramirez, Thomas C. Chasapis, Reza Loloee, Mercouri G. Kanatzidis, Alex Levchenko, and Stuart H. Tessmer, \textit{Scanning tunneling microscopy of superconducting topological surface states in Bi$_2$Se$_3$}, Phys. Rev. B 93 (2016) 220506(R).

\bibitem{Sedlmayr}
Nicholas Sedlmayr, E. W. Goodwin, Michael Gottschalk, Ian M. Dayton, Can Zhang, Erik Huemiller, Reza Loloee, Thomas C. Chasapis, Maryam Salehi, Nikesh Koirala, Mercouri G. Kanatzidis, Seongshik Oh, D. J. Van Harlingen, Alex Levchenko, and S. H. Tessmer, \textit{Dirac surface states in superconductors: a dual topological proximity effect}, preprint arXiv:1805.12330  [cond-mat.supr-con].

\bibitem{Lababidi}
Mahmoud Lababidi and Erhai Zhao, \textit{Microscopic simulation of superconductor/topological insulator proximity structures},  Phys. Rev. B \textbf{83} (2011) 184511.

\bibitem{Hugdal}
Henning G. Hugdal, Morten Amundsen, Jacob Linder, and Asle Sudb\o, \textit{Inverse proximity effect in s-wave and d-wave superconductors coupled to topological insulators}, Phys. Rev. B \textbf{99} (2019) 094505.

\bibitem{Bobkova}
I. V. Bobkova, and A. M. Bobkov, \textit{Electrically controllable spin filtering based on superconducting helical states}, Phys. Rev. B \textbf{96} (2017) 224505

\bibitem{Trang}
C. X. Trang, N. Shimamura, K. Nakayama, S. Souma, K. Sugawara, I. Watanabe, K. Yamauchi, T. Oguchi, K. Segawa, T. Takahashi, Yoichi Ando, and T. Sato, \textit{Conversion of a conventional superconductor into a topological superconductor by topological proximity effect},
Nature Comm. \textbf{11} (2020) 159.

\bibitem{Guigou}
M. Guigou, N. Sedlmayr, J. M. Aguiar-Hualde, and C. Bena, \textit{Signature of a topological phase transition in long SN junctions in the spin-polarized density of states}, Europhysics Letters \textbf{115} (2016) 47005.

\bibitem{Ptok}
Andrzej Ptok, Aksel Kobialka, and Tadeusz Doma\'nski, \textit{On controlling the bound states in quantum-dot hybrid-nanowire}, Phys. Rev. B \textbf{96} (2017) 195430.

\bibitem{Zhang-Zhang}
H. Zhang, C.-X. Liu, X.-L. Qi, X. Dai, Z. Fang, and S.-C. Zhang, \textit{Topological insulators in Bi$_2$Se$_3$, Bi$_2$Te$_3$, and Sb$_2$Te$_3$ with a single Dirac cone on the surface}, Nature Physics \textbf{5} (2009) 438.

\bibitem{Schnyder}
A. P. Schnyder, S. Ryu, A. Furusaki, and A. W. W. Ludwig, \textit{Classification of topological insulators and superconductors in three spatial dimensions}, Phys. Rev.
B \textbf{78} (2008) 195125.

\bibitem{Negele-Orland}
J. W. Negele and H. Orland, \textit{Quantum Many-Particle Systems},
Perseus Books (1998).

\end{thebibliography}

\end{document}